\documentclass[twocolumn,twoside]{IEEEtran}

\usepackage{epsfig}
\usepackage{amssymb}
\usepackage{amsxtra}
\usepackage{amsmath}
\usepackage{amsopn}
\usepackage{amsfonts}
\usepackage{latexsym}

\usepackage{theorem}

\usepackage{cite}

\title{Shortened Array Codes of Large Girth\\[0.60ex]}

\author{\large
Olgica Milenkovic,~\IEEEmembership{Member,~IEEE,}
Navin Kashyap,~\IEEEmembership{Member,~IEEE,}
and David Leyba
\vspace{-3.25ex}
\thanks{Manuscript submitted March 30, 2005, revised December 11, 2005.
This work NSF Grant CCF-0514857 awarded to O.\ Milenkovic,
a Lincoln Laboratory Fellowship awarded
to D.\ Leyba, and an NSERC Discovery Grant awarded to N.\ Kashyap.
Some of the results in this work were presented
at the 42nd Allerton Conference on Communication, Control and
Computing, Monticello, IL, USA, Sept.\ 2004.
}
\thanks{Olgica Milenkovic and David Leyba are with
the Department of Electrical and Computer Engineering,
University of Colorado, Boulder, CO 80309, USA.
(email: olgica.milenkovic@colorado.edu, david.leyba@colorado.edu)
}
\thanks{Navin Kashyap is with the Department of Mathematics
and Statistics at Queen's University, Kingston, ON K7L 3N6,
Canada. (email: {nkashyap@\ mast.queensu.ca}).
}
}

\renewcommand{\markboth}[2]
{\renewcommand{\leftmark}{#1}\renewcommand{\rightmark}{#2}}

\theoremstyle{plain}
\theorembodyfont{\normalfont\slshape}

\newtheorem{thm}{Theorem$\!$}
\newenvironment{theorem}
{\begin{thm}\hspace*{-1ex}{\bf.}}{\end{thm}}

\newtheorem{lem}[thm]{Lemma$\!$}
\newenvironment{lemma}{\begin{lem}\hspace*{-1ex}{\bf.}}{\end{lem}}

\newtheorem{prop}[thm]{Proposition$\!$}

\newtheorem{cor}[thm]{Corollary$\!$}
\newenvironment{corollary}{\begin{cor}\hspace*{-1ex}{\bf.}}{\end{cor}}

\newtheorem{defn}{Definition$\!$}
\newenvironment{definition}{\begin{defn}\hspace*{-1ex}{\bf.}}{\end{defn}}

\newcounter{enumrom}
\renewcommand{\theenumrom}{(\roman{enumrom})}

\makeatletter
\renewcommand{\@endtheorem}{\endtrivlist}
\makeatother

\makeatletter
\renewcommand{\thefigure}{{\bf \@arabic\c@figure}}
\renewcommand{\fnum@figure}{{\bf Fig.}\,\thefigure}
\makeatother

\renewcommand{\endproof}{~\QED\par\endtrivlist\unskip}

\def\Z{{\mathbb Z}}

\def\cC{{\cal C}}
\def\sfu{\textsf{u}}
\def\sfv{\textsf{v}}
\def\sfx{\textsf{x}}
\def\sfy{\textsf{y}}
\def\sfz{\textsf{z}}

\def\O{\Omega}

\begin{document}

\maketitle

\begin{abstract}
One approach to designing structured low-density parity-check (LDPC)
codes with large girth is to shorten codes with small girth in
such a manner that the deleted columns of the parity-check matrix
contain all the variables involved in short cycles. This approach is
especially effective if the parity-check matrix of a code is a
matrix composed of blocks of circulant permutation matrices, as is
the case for the class of codes known as array codes. We show how
to shorten array codes by deleting certain columns of their
parity-check matrices so as to increase their girth. The
shortening approach is based on the observation that for array
codes, and in fact for a slightly more general class of LDPC
codes, the cycles in the corresponding Tanner graph are governed
by certain homogeneous linear equations with integer coefficients.
Consequently, we can selectively eliminate cycles from an array code
by only retaining those columns from the parity-check matrix of the
original code that are indexed by integer sequences that do not
contain solutions to the equations governing those cycles.
We provide Ramsey-theoretic
estimates for the maximum number of columns that can be retained from the
original parity-check matrix with the property that the sequence
of their indices avoid solutions to various types of
cycle-governing equations. This translates to estimates of the rate
penalty incurred in shortening a code to eliminate cycles.
Simulation results show that for the
codes considered, shortening them to increase the girth can lead
to significant gains in signal-to-noise ratio in the case of
communication over an additive white Gaussian noise channel.
\end{abstract}


\section{Introduction}

Despite their excellent error-correcting properties,
low-density parity-check (LDPC) codes
with random-like structure \cite{G63},\cite[pp.\ 556--572]{Mac03}
have several shortcomings.
The most important of these is the lack of mathematical structure
in the parity-check matrices of such codes, which leads to
increased encoding complexity and prohibitively large storage
requirements. These issues can usually be resolved by using
structured LDPC codes, but at the cost of some performance loss.
This performance loss may be attributed to the fact that algebraic
code design techniques introduce various constraints on the set of
code parameters influencing the performance of belief propagation
decoding, so that it is hard to optimize the overall structure of
the code.

One parameter that is usually targeted for optimization
in the process of designing structured LDPC codes is the girth of
the underlying Tanner graph. Several classes of structured LDPC
codes with moderate and large values of girth and good performance
under iterative decoding are known, examples of which can be found
in \cite{GOS04,JW01,KPP04,KLF01,Mar82,MLZ04,RV01,VM04}.
In this paper, we focus our attention on a class of LDPC codes
termed array codes \cite{Fan00} (or equivalently, lattice codes
\cite{VM04}). These codes are quasi-cyclic, and have parity-check
matrices that are composed of circulant permutation matrices.
General forms of such parity-check matrices were investigated in
\cite{Fos04} and \cite{TSF01}, and codes of girth eight,
ten and twelve were obtained primarily through extensive computer
search.

Fossorier \cite{Fos04} considered a family of quasi-cyclic
LDPC codes closely related to array codes, and derived simple
necessary and sufficient conditions for such codes to have girth
larger than six or eight. Subsequently, codes with large girth
were constructed with the aid of computer search strategies which rely
on randomly generating integers until the conditions of the
theorem are met.

We generalize and extend the array code design methods in a
slightly different direction, and provide a less
computation-intensive approach to constructing codes with large
girth (including values exceeding eight). Our approach is based on
the observation that the existence of cycles in the Tanner graph
of an array code is governed by certain homogeneous linear equations.
We show that it is possible to exhaustively list all the equations
governing cycles of length six, eight and ten in an array code having
a parity-check matrix with a small number of ones in each column.
Thus, by shortening an array code in such a way as to only retain those
columns of its parity-check matrix whose indices form a sequence
that avoids solutions to some of these ``cycle-governing'' equations,
one can obtain array codes with a pre-specified distribution of cycles
of various lengths. This provides a means of studying the effects
of different classes of cycles on array code performance.
In particular, this technique can be used to entirely eliminate cycles
of short lengths, resulting in codes of girth up to twelve.
One special form of an array code of girth eight and
column-weight three was first described in \cite{VM04} and
\cite{BPI04}, where a good choice for the set of columns to be
retained from the original parity-check matrix was determined
using geometrical arguments.

Using techniques from graph theory and Ramsey theory, we provide
analytical estimates of the designed code rates achievable by
shortening an array code to improve girth, and present some useful
algorithms for identifying large sets of column-indices that avoid
solutions to cycle-governing equations. Simulation results show that
eliminating short cycles using this technique leads to significant
signal-to-noise ratio (SNR) gains, over the additive white Gaussian
nose (AWGN) channel. These codes also compare favorably with other
classes of structured LDPC codes in the literature, and in fact show
marked improvement in performance in some cases.

The remainder of the paper is organized as follows. Section~2
describes a generalization of the array code construction and
provides some definitions needed for the subsequent exposition. In
Section~3, we explicitly show how cycles in the Tanner graphs of
these codes are governed by certain homogeneous linear equations
with integer coefficients. We then go on to list the equations
governing cycles of length six, eight and ten in array codes
with parity-check matrices of small column-weight.
Section~4 contains bounds on the size
of the maximal sequence of column indices that contains no
solutions to certain homogeneous linear equations. A greedy
algorithm for constructing such sequences, as well as some simple
extensions thereof, are discussed in Section~5. Simulation results
are given in Section~6, with some concluding remarks presented in
Section~7. The proofs of some of the results of Section~4 are provided in
the Appendix.

\section{Array Codes}

Array codes \cite{Fan00} are structured LDPC codes with good performance
under iterative message-passing decoding.
Their parity-check matrix has the form
\begin{equation}
    H_{\mbox{\scriptsize arr}}=\left[\begin{array}{cccc}
      I & I & \cdots & I \\
      I & P & \cdots & P^{q-1} \\
     \cdots & \cdots & \cdots & \cdots \\
      I & P^{r-1} & \cdots & P^{(r-1)(q-1)} \\
    \end{array} \right],
\label{arrcode}
\end{equation}
where $q$ is an odd prime, $r$ is an integer\footnote{In this
paper, we will use the notation $[a,b]$ to denote the set $\{x \in
\Z: a \leq x \leq b\}$.} in $[1,q]$, $I$ is the $q \times q$
identity matrix, and $P$ is a $q \times q$ circulant permutation
matrix distinct from $I$. Recall that a permutation matrix is
a square matrix composed of 0's and 1's,
with a single 1 in each row and column.
A circulant permutation matrix is a permutation matrix that
is also circulant, \emph{i.e.}, the $i$th row of the matrix can be obtained
by cyclically shifting the $(i-1)$th row by one position to the right.
Typically, the matrix $P$ in (\ref{arrcode}) is chosen to be the matrix
\begin{equation}
P=\left[ \begin{array}{cccccc}
  0 & 1 & 0 & 0 & \ldots & 0 \\
  0 & 0 & 1 & 0 & \ldots & 0 \\
  \vdots & \vdots & \vdots & \vdots & \vdots & \vdots  \\
  0 & 0 & \ldots & 0 & 0 & 1\\
  1 & 0 & 0 & \ldots & 0 & 0\\
\end{array}\right]\,. \notag
\end{equation}
An LDPC code described by such a parity-check matrix is regular,
with length $q^2$ and co-dimension $r\,q$. The row and column
weights of such a code are $q$ and $r$, respectively.
Consequently, the rate $R$ of such codes is at least $1-r/q$.

We will consider the following more general form for a parity-check matrix:
\begin{equation}\label{generalized-array}
 H=\left[\begin{array}{cccc}
      P^{a_0 \cdot 0} & P^{a_0 \cdot 1} & \cdots & P^{a_0 \cdot (q-1)} \\
      P^{a_1 \cdot 0} & P^{a_1 \cdot 1} & \cdots & P^{a_1 \cdot (q-1)} \\
      \cdots & \cdots & \cdots & \cdots \\
      P^{a_{r-1} \cdot 0} & P^{a_{r-1} \cdot 1} & \cdots & P^{a_{r-1} \cdot (q-1)} \\
    \end{array} \right]
\end{equation}
where $a_0, a_1, \ldots, a_{r-1}$ is some sequence of $r$ distinct
integers from $[0,q-1]$. Each such parity-check matrix defines a
code. If the sequence $a_0, a_1, \ldots, a_{r-1}$ forms an
arithmetic progression (A.P.), {\em i.e.}, if there exists an
integer $a \neq 0$ such that $a_{i+1} - a_i = a$ for $i =
0,1,2,\ldots,r-2$, then we call the corresponding code a {\em
proper} array code (PAC). Note that if $a_0 = 0$, then the PAC
is simply an array code with parity-check matrix
$H_{\mbox{\scriptsize arr}}$ as in (\ref{arrcode}), since
the parity-check matrix in (\ref{generalized-array}) has
the same form as $H_{\mbox{\scriptsize arr}}$, as can be seen
by replacing $P$ in $H_{\mbox{\scriptsize arr}}$ by $P^a$.
If the sequence $a_0,a_1,\ldots,a_{r-1}$ does \emph{not} form
an A.P., then the corresponding code will
be referred to as an {\em improper} array code (IAC). The term
{\em array code} without further qualification will henceforth be
used to mean an IAC or a PAC.


Throughout the remainder of the paper, we will use the following
definitions/terminology:
\begin{itemize}
\item The odd prime $q$ used in defining the parity-check matrix of an
array code will be referred to as the \emph{modulus} of the code.
\item A {\em block-column (block-row)} of a parity-check matrix,
$H$, of an array code is the submatrix formed by
a column (row) of permutation matrices
from $H$. The $q$ block-columns of $H$ are indexed by the integers
from 0 to $q-1$, and the $r$ block-rows are indexed by the
integers from 0 to $r-1$. For example, the $j$th block-column of
$H$ is the matrix $[P^{a_0 \cdot j} \ P^{a_1 \cdot j}
\ P^{a_2 \cdot j} \ \ldots\ P^{a_{r-1} \cdot j}]^T$.
\item The term {\em block-row labels} will be
used to denote the integers in the sequence $a_0, a_1, \ldots,
a_{r-1}$ that define the matrix $H$ in (\ref{generalized-array}).
\item A {\em block-column-shortened array code}, or simply a {\em
shortened array code}, is a code whose parity-check matrix is
obtained by deleting a prescribed set of block-columns from the
parity-check matrix of an array code. \item The {\em labels} of the
block-columns retained in the parity-check matrix of the shortened code
are simply their indices in the parent code. For the parent code itself,
the terms ``label'' and ``index'' for a block-column can be used
interchangeably.
\item A {\em closed path} of length
$2k$ in any parity-check matrix of the form in
(\ref{generalized-array}) is a sequence of block-row and
block-column index pairs $(i_1,j_1)$, $(i_1,j_2)$, $(i_2,j_2)$,
$(i_2,j_3)$, \ldots, $(i_k,j_k)$, $(i_k,j_1)$, with $i_{\ell} \neq
i_{\ell+1}$, $j_{\ell} \neq j_{\ell+1}$, for $\ell =
1,2,\ldots,k-1$, and $i_k \neq i_1$, $j_k \neq j_1$.
\end{itemize}
The significance of closed paths arises from the following simple
but important result from \cite{Fan00} (see also
\cite[Theorem~2.1]{Fos04}):
\begin{theorem}
A cycle of length $2k$ exists in the Tanner graph
of an array code with parity-check matrix $H$ and block-row labels
$a_0,a_1,\ldots,a_{r-1}$ if and only if there exists a
closed path $(i_1,j_1)$, $(i_1,j_2)$,
$(i_2,j_2)$, $(i_2,j_3)$, \ldots, $(i_k,j_k)$, $(i_k,j_1)$ in $H$
such that
$$
P^{a_{i_1} \cdot j_1}\,(P^{a_{i_1} \cdot j_2})^{-1}\;P^{a_{i_2}
\cdot j_2}(P^{a_{i_2} \cdot j_3})^{-1} \;\cdots \;
P^{a_{i_k} \cdot j_k}(P^{a_{i_k} \cdot j_1})^{-1}
$$
evaluates to the identity matrix $I$.
\label{closed_path_theorem}
\end{theorem}
In fact, since $P$ is a $q \times q$ circulant permutation matrix, $P \neq I$,
and $q$ is prime, we can have $P^n = I$ if and only if
$n \equiv 0 \pmod{q}$. So, the
condition in the theorem is equivalent to
\begin{equation}
a_{i_1}(j_1 - j_2) + a_{i_2}(j_2 - j_3) + \cdots + a_{i_k} (j_k - j_1)
\equiv 0 \pmod{q},
\label{closed_path_eq1}
\end{equation}
which can also be written as
\begin{equation}
j_1(a_{i_1} - a_{i_k}) + j_2 (a_{i_2} - a_{i_1})
+ \cdots + j_k(a_{i_k} - a_{i_{k-1}}) \equiv 0 \pmod{q}.
\label{closed_path_eq2}
\end{equation}

Based on Theorem~\ref{closed_path_theorem}, it is easily seen
\cite{Fan00} that array codes are free of cycles of length four.
This is because a cycle of length four exists if and only if there
exist indices $i_1, i_2, j_1, j_2$, $i_1 \neq i_2$, $j_1 \neq j_2$
such that
$$
(a_{i_1} - a_{i_2})(j_1-j_2) \equiv 0 \pmod{q}.
$$
which is clearly impossible since $i_1 \neq i_2$ and $j_1 \neq j_2$.

On the other hand, an array code with a parity-check matrix of the form in
(\ref{arrcode}), with $q \geq 5, r \geq 3$, has cycles of length six.
An example is the closed path described by the coordinates
$(1,1),(1,2),(2,2)$, $(2,\frac{q+3}{2}),(0,\frac{q+3}{2}),(0,1)$,
which satisfies \eqref{closed_path_eq1}, since $a_i = i$ in this case, and
$$
1 \,  (1-2) + 2 \, (2-\frac{q+3}{2}) + 0 \, (\frac{q+3}{2} - 1)
= -q \equiv 0 \pmod{q}.
$$

In general, a closed path of length six in the parity-check matrix of an
array code must pass through three different block-rows, indexed
by $r_1,r_2,r_3$, and three different block-columns, indexed by $i,j,k$.
In the case of a PAC, the block-row labels $a_0,a_1,\ldots,a_{r-1}$
form an A.P.\ with common difference $a$, $0 < |a| < q$, and hence
\eqref{closed_path_eq2} reduces to
$$
a \, [i(r_1-r_3)+j(r_2-r_1)+k(r_3-r_2)] \equiv 0 \; \pmod  q.
$$
Thus, a PAC has a cycle of length six if and only if there exist distinct
block-row indices $r_1,r_2,r_3$ and distinct block-column indices $i,j,k$ such
that
\begin{equation}
\label{first-equation}
    i(r_1-r_3)+j(r_2-r_1)+k(r_3-r_2) \equiv 0 \; \pmod  q.
\end{equation}
Therefore, by shortening the
PAC so as to only retain block-columns with labels such that
(\ref{first-equation}) is never satisfied, we eliminate all cycles
of length six, obtaining a code of girth at least eight.

It is naturally of interest to extend this kind of analysis
to cover the case of cycles of length larger than six, and
utilize it to appropriately shorten an array code to
increase its girth.
The next section deals with the subject of identifying sequences
of block-column labels leading to codes with large girth.

\section{Array Codes of Girth Eight, Ten, and Twelve}

For clarity of exposition, in all subsequent derivations we will
focus only on the two special cases of array codes with column weight
three and four. The results presented can be extended in a
straightforward, albeit tedious, manner to codes with larger column weights.

\begin{theorem}
Let $\, \cC$ be a PAC with modulus $q$ whose parity-check matrix, $H$,
has column weight $r$. If $r=4$, then $\cC$ contains a cycle
of length six if and only if there exist three distinct block
columns in $H$ whose labels $i,j,k$ satisfy at least one of the
following two congruences:
\begin{equation}\label{girtheight}
    \begin{array}{c}
    -2i+j+k \equiv 0 \; \pmod q, \\
    -3i+j+2k \equiv 0 \; \pmod q.
    \end{array}
\end{equation}
If $r=3$, then $\cC$ contains a cycle of length six if and only if
there exist three distinct block columns whose labels $i,j,k$
satisfy the first of the two equalities. \label{sixcycle_theorem}
\end{theorem}

\begin{proof} The claim for $r=4$ follows immediately from
(\ref{first-equation}) once we note that any three block-row
indices $r_1,r_2,r_3 \in \{0,1,2,3\}$, $r_1 < r_2 < r_3$, must
satisfy one of the following: (i) $r_1-r_3=-2,\; r_3-r_2=1$, (ii)
$r_1-r_3=-3,\; r_3-r_2=1$, or (iii) $r_1-r_3=-3,\; r_3-r_2=2$.

The proof for the $r=3$ case similarly follows from the fact that
the only possible choice for the set of three distinct block-row
labels in this case is $\{0,1,2\}$.
\end{proof} \mbox{ }

A useful consequence of the above result is Corollary~\ref{r3_cor}
below, to state which it is convenient to introduce the following definition.
Here, and in the rest of the paper, the set of positive integers
is denoted by $\Z^+$, and given an $N \in \Z^+$, the ring
of integers modulo $N$ is denoted by $\Z_N$.

\begin{definition} 
A sequence of distinct non-negative integers $n_1,n_2,n_3,\ldots$
is defined to be a \emph{non-averaging sequence}
if it contains no term that is the
average of two others, {\em i.e.}, $n_i+n_j=2n_k$ only if $i=j=k$.
Similarly, given an $N \in \Z^+$, a sequence of distinct
integers $n_1,n_2,n_3,\ldots$ in $[0,N-1]$ is non-averaging
over $\Z_N$ if $n_i+n_j \equiv 2n_k \pmod{N}$ implies that $i=j=k$.
\label{non-averaging_def}
\end{definition}

It is clear from the definition that a sequence is non-averaging
if and only if it contains no non-constant three-term A.P.
The following result is a simple consequence of
Theorem~\ref{sixcycle_theorem} and
Definition~\ref{non-averaging_def}.

\begin{corollary}
Let $H$ be the parity-check matrix of a PAC with modulus $q$,
consisting of three block-rows, and let $A$ be the $3q \times mq$
matrix obtained by deleting some $q-m$ block-columns from $H$.
The shortened array code with parity-check matrix $A$ has girth at least eight
if and only if the sequence of labels of the block-columns in $A$ forms a
non-averaging sequence over $\Z_q$.
\label{r3_cor}
\end{corollary}

To extend the above result to PAC's with four block-rows, we require
the following generalization of Definition~\ref{non-averaging_def}.

\begin{definition} Let $c$ be a fixed
positive integer. A sequence of distinct non-negative integers
$n_1,n_2,n_3,\ldots$ is defined to be a 
\emph{$c$-non-averaging sequence} if $n_i+c n_j=(c+1)n_k$ implies
that $i=j=k$. We extend this definition as before to sequences
over $\Z_N$, for an arbitrary $N \in \Z^+$.
\label{c-non-averaging_def}
\end{definition}

Note that a sequence is $c$-non-averaging if and only if it
does not contain three elements of the form
$n,n+t,n+(c+1)t$,
for some integers $n,t$, with $t > 0$. We can now state the following
corollary to Theorem~\ref{sixcycle_theorem}.

\begin{corollary}
Let $H$ be the parity-check matrix of a PAC with modulus $q$,
consisting of four block-rows, and let $A$ be the $4q \times mq$ matrix
obtained by deleting some $q-m$ block-columns from $H$.
The shortened array code with parity-check matrix $A$ has girth
at least eight if and only if the sequence of block-column labels in $A$
is non-averaging and  2-non-averaging over $\Z_q$.

\label{r4_cor}
\end{corollary}

We next consider the case of cycles of length eight. By the reasoning
used to derive (\ref{first-equation}), it follows from
Theorem~\ref{closed_path_theorem} that a PAC contains a
cycle of length eight if and only if its parity-check matrix contains a closed
path of the form $(r_1,i),(r_1,j),(r_2,j),(r_2,k),(r_3,k),(r_3,l),(r_4,l),(r_4,i)$
such that
\begin{equation}\label{eightcycle_eq}
i(r_1-r_4)+j(r_2-r_1)+k(r_3-r_2)+l(r_4-r_3) \equiv 0\, \pmod q
\end{equation}
Note that closed paths of length eight may pass through two, three or
four different block-columns of the parity-check matrix of the PAC.

Let us first consider the situation where a closed path passes
through exactly two different block-columns. Let $i$ and $j$ be
the labels of these block-columns. This closed path forms a cycle
of length eight if and only if (\ref{eightcycle_eq}) is satisfied
with $k=i$ and $l=j$. A re-grouping of terms results in the
equation
\begin{equation}
(i-j)(r_1 + r_3 - r_2 - r_4) \equiv 0 \, \pmod q \notag
\end{equation}
which, for $i \neq j$, is satisfied if and only if
\begin{equation}\label{two_col_eightcycle_eq}
r_1 + r_3 - r_2 - r_4 \equiv 0 \, \pmod q.
\end{equation}
Now, observe that for a PAC with column-weight $r \geq 3$, the
above equation is always satisfied by taking
$r_1 = 0$, $r_2 = 1$, $r_3 = 2$ and $r_4 = 1$. This shows that in
a PAC with column-weight $r \geq 3$, {\em any} pair of
block-columns is involved in a cycle of length eight. Hence,
shortening will never be able to eliminate cycles of length eight
from such a PAC (except obviously in the trivial case where we
delete all but one block-column), implying that shortened PAC's
can have girth at most eight. We record this fact in the lemma below.

\begin{lemma}
A shortened PAC of column-weight at least three has girth at most eight.
\label{PAC_girth8_lemma}
\end{lemma}

The following theorem provides the constraining equations that govern
cycles of length eight involving three
of four different block-columns in a PAC with row-weight $q$ and
column-weight three or four. The proof is along the lines of that of
Theorem~\ref{sixcycle_theorem}, and is omitted.

\begin{theorem} \label{eightcycle_theorem}
In a PAC with modulus $q$ and column-weight $r=3$, the constraining equations,
over the ring $\Z_q$, for the block-column labels $i,j,k,l$ specifying
cycles of length eight involving three or four different block-columns are
\begin{equation}\label{r3_girthten}
\begin{array}{cc}
    i-j-k+l=0, & 2i-j-2k+l=0 \\
    2i+j-3k=0, & 2i-j-k=0
\end{array}
\end{equation}
For PAC's with modulus $q$ and column-weight $r=4$, the set of constraining
equations, over $\Z_q$, for the labels $i,j,k,l$ that describe cycles of
length eight involving three or four different block-columns is
\begin{eqnarray}\label{r4_girthten}
\begin{array}{c}
\begin{array}{cc}
    3i-j-k-l=0, & 3i-2j-2k+l=0, \\
    3i-3j+k-l=0, & 3i-3j+2k-2l=0, \\
    2i-2j+k-l=0, & i+j-k-l=0, \\
\end{array} \\
 2i-j-k=0,\ \ \ 4i-3j-k=0,\ \ \ 3i-2j-k=0
\end{array}
\end{eqnarray}
\end{theorem}

\begin{figure}[t]
\centering \epsfig{file=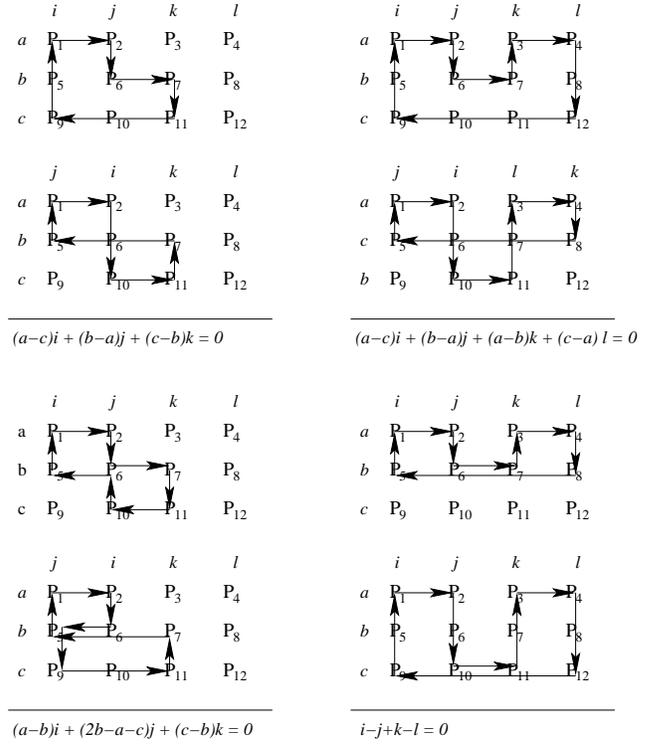, width=8.5cm}
\caption{Some cycles of lengths six and eight, and their governing equations.}
\label{cvseq}
\end{figure}

Figure~\ref{cvseq} shows the structures of some cycles of lengths six and eight,
and provides the modulo-$q$ equation governing each such cycle. The
generic variables $a,b,c$ and $i,j,k,l$ represent the
block-row and block-column labels, respectively. The equations governing all
such cycles are also summarized in Tables~\ref{PACr3_table} and \ref{PACr4_table}.

It should be abundantly clear by now that we can eliminate a
large number of cycles of length eight from a PAC by selectively deleting
some of its block-columns, retaining only those block-columns the
set of whose labels does not contain solutions to some or all of
the equations listed in Theorem~\ref{eightcycle_theorem}.
Note also that the equations listed
in \eqref{girtheight}, upon relabeling the variables if necessary,
form a subset of the equations listed in \eqref{r3_girthten}, as well as
of those in \eqref{r4_girthten}. Hence, if we shorten a PAC
in such a way as to retain only those block-columns whose labels form a
non-averaging and 2-non-averaging sequence, not only does the resultant
shortened code have no cycles of length six, but it also has
fewer cycles of length eight than the original code.


As observed earlier, shortened PAC's cannot have girth larger than
eight. This is a direct consequence of the fact that the block-row
labels of a shortened PAC with column-weight at least three always contain
a solution to \eqref{two_col_eightcycle_eq}, and hence any such code always
contains cycles of length eight that pass through pairs of
distinct block-columns. On the other hand, IAC's can be constructed in
such a way as to avoid cycles of length eight that involve
only two different block-columns. Analogous to \eqref{two_col_eightcycle_eq},
the equation governing such cycles in an IAC is
\begin{equation}\label{two_col_eightcycle_eq_iac}
a_{r_1} + a_{r_3} - a_{r_2} - a_{r_4} \equiv 0 \, \pmod q.
\end{equation}
Thus, if the block-row labels of the IAC are chosen so that
they do not contain solutions to \eqref{two_col_eightcycle_eq_iac},
then such eight-cycles cannot arise. Examples of such sets
of block-row labels are $\{{0,1,3\}}$ for an IAC with three block-rows, and
$\{{0,1,3,7\}}$ for an IAC with four block-rows. Such IAC's can be shortened
to yield codes with girth ten or twelve, provided that the block-column
labels retained in the shortened code avoid a set of constraining equations
analogous to \eqref{girtheight}, \eqref{r3_girthten} and \eqref{r4_girthten}.
The equations governing cycles of lengths six, eight and ten for IAC's with
three block-rows ($r=3$) and label set $\{0,1,3\}$ are listed in
Table~\ref{IACr3_table}. Similarly, Table~\ref{IACr4_table} lists the
twenty-eight equations governing cycles of lengths six and eight in IAC's
with four block-rows ($r=4$) and label set $\{0,1,3,7\}$. There are
more than fifty equations governing cycles of length ten in IAC's with $r=4$.
These equations were obtained via an exhaustive computer-aided analysis of
all the possible structures that cycles can have in these codes.

It is worth pointing out that Tables~\ref{PACr3_table}--\ref{IACr4_table}
need not only be used to construct codes with a prescribed girth, but can also
be used to design codes with a pre-specified set of cycles. This can
help in studying the effects of various cycle classes on the performance
of a code.

The structure of the parity-check matrix in an array code allows us to
use existing results in the literature to obtain upper and lower bounds
on the minimum distance, $d$, of such codes. A lower bound on $d$ for
regular LDPC codes was derived in \cite{Tan81}:
\begin{equation}
d \geq \left\{\begin{array}{c}
  2\; \frac{(r-1)^{(g-2)/4}-1}{r-2}+\frac{2}{r}\,(r-1)^{(g-2)/4},\;\; g/2 \;\; \text{odd} \\
  \;\;\;\;\;\;\;\;\;\;\;2\; \frac{(r-1)^{g/4}-1}{r-2}, \; \;\;\;\;\;\;\;\;\;
  \;\;\;\;\;\;\;\;\;\;\;\;\;\;g/2 \;\; \text{even}\\
\end{array} \right.
\end{equation}
where $g$ is the girth of the code and $r$ is the column-weight of
the parity-check matrix (\emph{i.e.}, the degree of any variable
node). This bound can be improved slightly in some cases by noting
that the minimum distance of an array code must be even,
since the code can have even-weight codewords only. This is
a consequence of the fact that within any block-row, $[P^{a_i
\cdot 0} \ P^{a_i \cdot 1} \ P^{a_i \cdot 2} \ \ldots\ P^{a_i
\cdot (q-1)}]$, of the parity check matrix of an array code, the
rows sum to $[1 \ 1 \ 1 \ \ldots \ 1]$, and hence the dual of an
array code always contains the all-ones codeword.

For bounding $d$ from above, we make use of a particularly
elegant result due to MacKay and Davey \cite{MD99}, which shows
that parity-check matrices containing an $r \times (r+1)$ grid of
permutation matrices $P_{i,j}$ that commute (\emph{i.e.}, for which
$P_{i,j}P_{k,l}=P_{k,l}P_{i,j}$) must have minimum distance at
most $(r+1)!$.  Table~\ref{dmin_table} lists the lower and upper bounds
on minimum distance for array codes with column-weight $r \in \{3,4\}$
and girth $g \in \{8,10,12\}$.

\begin{table*}
\caption{Bounds on the minimum distance, $d$, of array codes
for various values of column-weight, $r$, and girth, $g$.}
\label{dmin_table}
\centering
\begin{tabular}{c|c|c|c|c|}
\cline{2-5}
& $r=3$ & $r=3$ & $r=4$ & $r=4$ \\ \hline
\multicolumn{1}{|c|}{Girth $g$} &
Lower bound on $d$ & Upper bound on $d$ & Lower bound on $d$ & Upper bound on $d$\\
\hline \multicolumn{1}{|c|}{8} & 6 & 24 & 8 & 120\\ \hline
\multicolumn{1}{|c|}{10} & 10 & 24 & 14 & 120 \\ \hline
\multicolumn{1}{|c|}{12} & 14 & 24 & 26 & 120 \\ \hline
\end{tabular}
\end{table*}

\subsection{The Code Mask}

\begin{figure}
\centering \epsfig{file=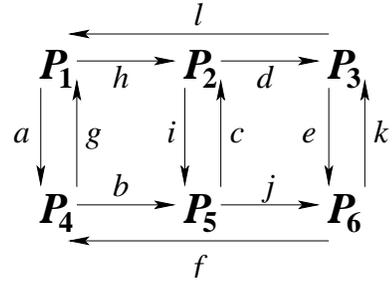, width=5cm}
   \caption{\label{12-cycle} Cycle of length twelve in an array code.}
\end{figure}

Array codes, as well as the general class of quasi-cyclic LDPC
codes with parity-check matrices consisting of blocks of circulant
permutation matrices, cannot have girth exceeding twelve
\cite{Fos04}. This is most easily seen by examining the
example in Figure~\ref{12-cycle}. There, a sub-matrix of a parity
check matrix consisting of circulant permutation blocks $P_i$,
$i=1,2,\ldots,6$, is shown, along with a directed closed path
labeled $abcdefghijkl$ that traverses the blocks. Setting
$P_i = P^{b_i}$ for some circulant permutation matrix $P$ and
exponents $b_i$, we see that the condition in
Theorem~\ref{closed_path_theorem} is satisfied, since
\begin{equation} \label{twzero}
b_1-b_4+b_5-b_2+b_3-b_6+b_4-b_1+b_2-b_5+b_6-b_3=0.
\end{equation}
Thus, length-12 cycles are guaranteed to exist in any quasi-cyclic LDPC
code with parity-check matrix consisting of blocks of circulant
permutation matrices.

Nevertheless, using a \emph{masking approach}, array codes can be
modified so that their girth exceeds twelve. Masks were
introduced in \cite{MPV03} for the purpose of increasing
the girth of codes as well as for constructing irregular LDPC
codes. As an illustrative example, consider the matrix $M$ in \eqref{M_eq}
below. It consists of $q \times q$ zero matrices \textbf{0} and $q
\times q$ circulant permutation blocks $P_i=P^{b_i}$, for some
integers $b_i$. One can view $M$ as arising from a parity-check
matrix of an array code, or more generally, a
quasi-cyclic code with circulant permutations blocks,
from which some blocks are ``zeroed out'' according to a given
mask. The matrix $M$ does not contain a submatrix of the form
depicted in Figure~\ref{12-cycle}. Consequently, there exist no
length-12 cycles that traverse exactly six permutation matrix blocks.
Of course, this is achieved at the expense of increased code length
(for the given example, the length has to be doubled).
Other kinds of length-12 cycles may still exist, but these are governed
by non-trivial homogeneous linear equations similar in form to
those governing shorter cycles, and can be eliminated by a judicious
choice of the exponents $b_i$.

\begin{equation}
M=\left(%
\begin{array}{cccccccc}
  P_1 & P_2 & P_3 & \textbf{0} & \textbf{0} & \textbf{0} & P_4 & \textbf{0} \\
  \textbf{0} & \textbf{0} & \textbf{0} & P_5 & P_6 & P_7 & \textbf{0} & P_8 \\
  P_9 & P_{10} & \textbf{0} & P_{11} & P_{12} & \textbf{0} & \textbf{0} & \textbf{0} \\
  \textbf{0} & \textbf{0} & P_{13} & \textbf{0} & \textbf{0} & P_{14} & P_{15} & P_{16} \\
  \textbf{0} & P_{17} & P_{18} & P_{19} & \textbf{0} & P_{20} & \textbf{0} & \textbf{0} \\
  P_{21} & \textbf{0} & \textbf{0} & \textbf{0} & P_{22} & \textbf{0} & P_{23} & P_{24} \\
\end{array}%
\right)
\label{M_eq}
\end{equation}

\section{Avoiding Solutions to Cycle-Governing Equations}

Cycle-governing equations, such as those listed in
Tables~\ref{PACr3_table}--\ref{IACr4_table}, are always of the
following type:
\begin{equation}\label{part_reg_eq}
\sum_{i=1}^m c_i \sfu_i \equiv 0 \; \pmod{q},
\end{equation}
the integer $m$ being the number of distinct
block-columns through which the cycle passes,
the $\sfu_i$'s being variables\footnote{To avoid the sloppiness of
using $u_i$ to denote both a variable and a value it can take, we will
make typographical distinctions between the two whenever necessary.}
that denote the labels of those $m$ block-columns,
and the $c_i$'s being fixed nonzero integers
(independent of $q$) such that $\sum_{i=1}^m c_i = 0$.
This is because all such equations arise as special
cases of an equation of the form \eqref{closed_path_eq2},
and clearly, $(a_{i_1} - a_{i_k}) + (a_{i_2} - a_{i_1}) + \cdots +
(a_{i_k} - a_{i_{k-1}}) = 0$. Any solution
$\mathbf{u} = (u_1,u_2,\ldots,u_m)$ to \eqref{part_reg_eq},
with $u_i \in [0,q-1]$, and such that
$u_i \neq u_j$ when $i \neq j$, represents a cycle passing
through the $m$ block-columns whose labels form the solution
vector $\mathbf{u}$.

To avoid potential ambiguity, we establish some terminology that we
will use consistently in the rest of the paper.
Given a homogeneous linear equation of the form
$\sum_{i=1}^m c_i \sfu_i = 0$, we refer
to a vector $(u_1,u_2,\ldots,u_m) \in {[0,q-1]}^{m}$ as a
\emph{solution over $\Z_q$} to the equation if
$\sum_{i=1}^m \, c_i u_i \equiv 0 \pmod{q}$.
If $\mathbf{u} = (u_1,u_2,\ldots,u_m) \in {\Z}^{m}$ is such that
$\sum_{i=1}^m \, c_i u_i = 0$, then $\mathbf{u}$
is referred to as an \emph{integer solution} to the equation.
In both cases, a solution $\mathbf{u} = (u_1, u_2, \ldots, u_m)$
to \eqref{part_reg_eq}, with all the $u_i$'s distinct,
will be referred to as a \emph{proper} solution.

The design of a shortened array code typically involves determining
the smallest prime $q$ for which there exists a sequence of integers
$S \subset [0,q-1]$ of some desired cardinality $s$, such that there
is no proper solution with entries in $S$ to any equation within a
certain set of cycle-governing equations. This choice of $q$ would
guarantee the smallest possible code length, equal to $q\,s$, for a
PAC or an IAC with prescribed girth, column-weight $r$ and designed
code rate $R = 1-r/s$. For example, if we seek an IAC with $r=3$,
designed rate $R=1/2$ and girth ten, then we need the smallest $q$
that guarantees the existence of a set $S$ of cardinality at least
six that does not contain a proper solution to any of the equations
listed in Table~\ref{IACr3_table}. It is therefore useful to
estimate, as a function of $q$, the size of the largest subset of
$[0,q-1]$ that avoids proper solutions to certain linear equations
of the form given in \eqref{part_reg_eq}. In this section, we
provide a number of results that bound the size of such a largest
subset.

Equations of the form $\sum_{i=1}^{m} c_i \sfu_i = 0$, with
$\sum_{i=1}^m c_i = 0$, have been extensively studied in Ramsey
theory \cite[Chapter~3]{GRS90}, \cite[Chapter 9]{LR04}.
It is known \cite[Fact~3]{FGR88} that any such equation that is
not of the form $\sfu_1 - \sfu_2 = 0$ (or an integer multiple of it)
has a proper solution. In fact \cite[Theorem~2]{FGR88},
for any $\epsilon > 0$ and sufficiently large $N$,
if $L \subset [1,N]$ is such that $|L| \geq \epsilon N$,
then $L$ contains a proper solution to such an equation.
This implies the following result:

\begin{theorem} \label{th-dens}
Let $m \geq 3$, and let $c_i$, $i = 1,2,\ldots,m$, be nonzero
integers such that $\sum_{i=1}^m c_i = 0$. For an arbitrary $q
\geq 1$, let $s(q)$ be the size of the largest subset of $[0,q-1]$
that does not contain a proper solution to
$\sum_{i=1}^m c_i \sfu_i \equiv 0 \pmod{q}$. Then,
$$\lim_{q\rightarrow\infty} \frac{s(q)}{q} = 0.$$
\end{theorem}
\begin{proof} Let $S(q) \subset [0,q-1]$
be a set of size $s(q)$ that does not contain any proper solution
to $\sum_{i=1}^m c_i \sfu_i \equiv 0 \pmod{q}$. Clearly, $S(q)$ does not
contain a proper solution to $\sum_{i=1}^m c_i \sfu_i = 0$ (without
the modulo-$q$ reduction) as well. Note that since
$(1,1,\ldots,1)$ is a solution to $\sum_{i=1}^m c_i \sfu_i = 0$,
$(u_i)$ is a solution iff $(u_i+1)$ is a solution. Thus, $L(q) =
S(q) + 1 = \{j + 1: j \in S(q)\}$ is a set of cardinality $s(q)$
in $[1,q]$ that does not contain a proper solution to
$\sum_{i=1}^m c_i \sfu_i = 0$. Hence, for any $\epsilon > 0$, we must
have $s(q) < \epsilon q$ for all sufficiently large $q$, and the
desired result follows.
\end{proof} \mbox{}

We have thus established that the size of a subset of $[0,q-1]$
containing no proper solution to any equation from a given set of
cycle-governing equations grows sub-linearly in $q$. This is a
disappointing result from the point of view of our strategy of
shortening array codes to eliminate cycles. Indeed, starting with an
array code of column-weight $r$, length $q^2$ and designed rate
$1-r/q$, if we shorten the code so as to eliminate cycles governed
by an equation of the form $\sum_{i=1}^m c_i \sfu_i \equiv 0
\pmod{q}$, the resulting shortened code can have rate no larger than
$1 - r/s(q)$, where $s(q)$ is as defined in the statement of
Theorem~\ref{th-dens}. Since $s(q)/q$ goes to 0 as $q$ increases,
the rate penalty associated with shortening is severe for large
values of $q$ (or equivalently, for large values of the length of
the parent code). However, from a practical standpoint, this does
not appear to be a problem, as for the moderate values of $q$ useful
in practical code constructions, the rate penalty incurred by
shortening remains within reasonable limits. Consequently, it is
possible to construct, for example, designed rate-1/2 codes of girth
eight and ten that perform much better than the comparable codes in
the existing literature, as we shall see in Section~6.

A precise estimate of the rate at which $s(q)/q$ goes
to zero for various types of cycle-governing equations can be
very useful for the purpose of practical code design, as this
provides us with an understanding of how the rate penalty incurred
in shortening an array code changes with the modulus $q$.
More generally, given a collection, $\O$, of homogeneous linear
equations over $\Z_q$ of the form~\eqref{part_reg_eq}, let $s(q;\O)$
be the size of the largest subset of $[0,q-1]$ that does not contain
a proper solution over $\Z_q$ to \emph{any} of the equations in $\O$.
From the result of Theorem~\ref{th-dens}, it is clear that
$s(q;\O)$ grows sub-linearly with $q$. In the rest of this section,
we provide upper and lower bounds on $s(q;\O)$ for various choices of
$\O$.

\subsection{Upper bounds on $s(q;\O)$}

Explicit upper bounds for $s(q;\O)$ can be obtained for any $\O$ containing
an equation (over $\Z_q$) of the form $2\sfx - \sfy - \sfz = 0$ or
$\sfx + \sfy - \sfz - \sfu = 0$. These equations have been extensively studied
in other contexts, and in such cases, there are good estimates
available for the growth rate of sequences avoiding solutions to
these equations.

Recall from Definition~\ref{non-averaging_def} that sequences
avoiding proper solutions to $2\sfx - \sfy - \sfz = 0$ are called
non-averaging sequences. Correspondingly, sequences avoiding
proper solutions to the equation $\sfx+\sfy-\sfz-\sfu = 0$ are called
Sidon sequences (see \emph{e.g.}\ \cite{OBr02}),
as made precise by the definition below.

\begin{definition} A \emph{Sidon sequence} is a sequence of distinct
integers $n_1,n_2,n_3,\ldots$ with the property that
for all $i,j,k,l$ such that $i \neq j$, $k \neq l$, $n_i + n_j = n_k + n_l$
if and only if $\{i,j\} = \{k,l\}$. Similarly, given an $N \in \Z^+$,
a \emph{Sidon sequence over $\Z_N$} is a sequence of distinct integers
$n_1,n_2,n_3,\ldots$ in $[0,N-1]$ such that for all
$i,j,k,l$ with $i \neq j$, $k \neq l$,
$n_i + n_j = n_k + n_l \pmod{N}$ if and only if $\{i,j\} = \{k,l\}$.
\end{definition}

Upper bounds on the sizes of non-averaging sequences and Sidon sequences
over $\Z_N$ are given in the next lemma. Observe that for any $N \in \Z^+$,
a non-averaging sequence over $\Z_N$ is automatically a
non-averaging sequence (over $\Z^+$). The result of part (a) of the
lemma is thus a straightforward application of the classical upper bound,
due to Roth \cite[Section 4.3, Theorem~8]{GRS90}, on
the cardinality of the largest non-averaging sequence in
$[0,N-1]$.

\begin{lemma} (a) (Roth's theorem) The cardinality of any
non-averaging sequence over $\Z_N$ is bounded from above
by $c_0 N/\log\log\, N$, for some fixed constant $c_0 > 0$.

(b) For any odd integer $N > 0$, the cardinality of a Sidon sequence
over $\Z_N$ is bounded from above by $\sqrt{N-3/4} + 1/2$.
\label{roth_sidon_lemma}
\end{lemma}

We defer the proof of part (b) of the above lemma to the Appendix.
In terms of the quantity $s(q;\O)$, the lemma can be re-stated as:
\begin{itemize}
\item[(a)] If $\O$ contains the equation $2\sfx-\sfy-\sfz
\equiv 0 \pmod{q}$, then $s(q;\O) \leq c_0 \, q/\log\log\, q$,
for some fixed constant $c_0 > 0$.
\item[(b)] If $\O$ contains the equation $\sfx+\sfy-\sfz-\sfu
\equiv 0 \pmod{q}$, then $s(q;\O) \leq \sqrt{q-3/4} + 1/2$.
\end{itemize}

In a PAC with modulus $q$ and column-weight $r \geq 3$, the equation
$2\sfx - \sfy - \sfz \equiv 0 \pmod{q}$ always governs six-cycles,
as can be seen by setting $r_1 = 0$, $r_2 = 1$ and $r_2 = 2$ in
\eqref{first-equation}. So, if a shortened PAC has girth eight,
then its sequence of block-column labels must not contain
solutions to $2\sfx - \sfy - \sfz \equiv 0 \pmod{q}$,
\emph{i.e.}, must be non-averaging over $\Z_q$.
Hence by Lemma~\ref{roth_sidon_lemma}(a), the number of block-columns
in the parity-check matrix
of the shortened PAC cannot exceed $c_0 \, q/\log\log q$.

Similarly, in an array code with modulus $q$,
the equation $\sfx+\sfy-\sfz-\sfu \equiv 0 \pmod{q}$
always governs eight-cycles that pass through any two
distinct block-rows and four distinct block-columns
(see, for example, the cycles on the bottom-right of Figure~\ref{cvseq}).
So, if an array code is shortened to obtain
girth ten, then the sequence of block-column labels
retained in the shortened code must be a Sidon sequence over $\Z_q$,
and therefore, Lemma~\ref{roth_sidon_lemma}(b) applies.
We have thus proved the following theorem.

\begin{theorem}
(a) The number of block-columns in the parity-check matrix
of a shortened PAC with modulus $q$, column-weight $r \geq 3$
and girth eight cannot exceed $c_0 \, q/\log\log\, q$.

(b) The number of block-columns in the parity-check matrix of
a shortened array code with modulus $q$, column-weight $r \geq 2$
and girth ten is at most $\sqrt{q-3/4} + 1/2$.
\label{block-col_theorem}
\end{theorem}

Roughly speaking, the above theorem says that the rate of
a shortened PAC with modulus $q$, column-weight $r \geq 3$
and girth eight cannot be more than $1 -\frac{\log\log q}{c_0q} \, r$.
Similarly, the rate of a shortened array code with modulus $q$,
column-weight $r \geq 2$ and girth ten is, as a rough estimate, bounded
from above by $1 - \frac{r}{\sqrt{q}}$.

It is natural to want to compare the bounds of Theorem~\ref{block-col_theorem}
to those obtained from the application of the Moore bound
to the Tanner graphs of array codes. The Moore bound\footnote{To be
correct, this should be called a Moore-type bound, as the original Moore
bound (see \cite[p.\ 180]{Big93}) only applies to regular graphs.} for a
bipartite graph \cite{Hoo02} bounds the number of vertices
in the graph in terms of the girth and the average left and right degrees.
Consider a bipartite graph with $n_L$ left vertices, $n_R$ right vertices,
$m$ edges and girth $g$. Let $d_L = \frac{m}{n_L}$ be the average left degree,
$d_R = \frac{m}{n_R}$ the average right degree. Then,

\begin{eqnarray}
n_L &\geq& \sum_{i=0}^{g/2-1}\, (d_R-1)^{\lceil\,i/2\rceil}\,
(d_L-1)^{\lfloor\,i/2\rfloor} \label{nL_bnd} \\
n_R &\geq& \sum_{i=0}^{g/2-1}\, (d_L-1)^{\lceil\,i/2\rceil}\,
(d_R-1)^{\lfloor\,i/2\rfloor}. \label{nR_bnd}
\end{eqnarray}
The above bounds are easily proved for bi-regular bipartite graphs,
\emph{i.e.}, graphs in which each left (resp.\ right)
vertex has degree $d_L$ (resp.\ $d_R$).

Now, the Tanner graph of an array code of modulus $q$, column-weight $r$
and having $s$ block-columns is bi-regular with $n_L = qs$, $n_R = qr$,
$d_L = r$ and $d_R = s$. So, for such a Tanner graph of girth eight,
the bound in \eqref{nL_bnd} becomes
\begin{eqnarray*}
q s &\geq& 1 + (s-1) + (s-1)(r-1) + (s-1)^2 (r-1) \\
&=& s \, [1 + (s-1)(r-1)],
\end{eqnarray*}
which yields the bound
\begin{equation}
s \leq 1 + \frac{q-1}{r-1}.
\label{g8_bnd}
\end{equation}
The bound in \eqref{nR_bnd} also gives exactly the same result.
Note that this bound is, asymptotically in $q$, looser than the bound
in Theorem~\ref{block-col_theorem}(a). But for practical purposes,
this is a more useful bound than that of the theorem because the $c_0$ in
the theorem is not explicitly specified.

On the other hand, applying \eqref{nL_bnd} to the Tanner graph of an
array code of girth ten, we get
$$
qs \geq s \, [1 + (s-1)(r-1)] + (s-1)^2(r-1)^2,
$$
which upon re-arrangement becomes
$$
r(r-1)s^2 - [r(2r-3)+q] \, s + (r - 1)^2 \leq 0.
$$
Solving for $s$ now yields
\begin{equation}
s \leq \frac{q + r(2r-3) + \sqrt{(q+r(2r-3))^2 - 4r(r-1)^3}}{2r(r-1)}.
\label{g10_bnd}
\end{equation}
For $q \gg r^2$, this upper bound is roughly $\frac{q}{r(r-1)}$. It is clear
that in most cases of interest, this is not as good a bound as that
of Theorem~\ref{block-col_theorem}(b). We would like to remark that
another upper bound can be obtained via \eqref{nR_bnd},
but this turns out to be looser than the bound in \eqref{g10_bnd}.

We summarize the above bounds in the following theorem.

\begin{theorem}
(a) The number of block-columns in the parity-check matrix
of a shortened array code with modulus $q$, column-weight $r$
and girth eight cannot exceed $1 + (q-1)/(r-1)$.

(b) The number of block-columns in the parity-check matrix of
a shortened array code with modulus $q$, column-weight $r$
and girth ten is at most
$$
\frac{q + r(2r-3) + \sqrt{(q+r(2r-3))^2 - 4r(r-1)^3}}{2r(r-1)}.
$$
\label{moore_theorem}
\end{theorem}

\subsection{Lower bounds on $s(q;\O)$}

We next consider the converse problem of finding lower
bounds on the size of integer sequences avoiding solutions to a collection
of cycle-governing equations. The problem of constructing long sequences
of integers that do not contain solutions to certain kinds of homogeneous
linear equations has a long history. For example, large non-averaging
subsets of $[1,N]$ were described or constructed by Behrend
\cite{Beh46}, Moser \cite{Mos53} and Rankin \cite{Ran62}, using
geometrical arguments. We will generalize some of
these results to cover certain classes of equations of the form given in
\eqref{part_reg_eq}.

We start with a lower bound on the maximum length of sequences that
are $c_i$-non-averaging over $\Z_q$, for $\ell$
distinct integers $c_i \in [2,q-2]$. The proof of this bound is
provided in the Appendix.

\begin{theorem} Let $\ell \geq 1$, and let $\O$ be the collection of equations
$$
\sfx + c_i\sfy = (c_i+1)\sfz, \ \ \ i = 1,2,\ldots,\ell,
$$
for some constants $c_i \in [1,q-2]$ such that $c_i \neq c_j$ for $i \neq j$.
Then,
$$
s(q;\O) \geq \left(\frac{3q^2}{\ell(q-1)}\right)^{1/3}.
$$
\label{c-non-av_theorem}
\end{theorem}


The lower bound derived in the theorem above is quite loose. For
example, for $q=241$, a greedy algorithm (to be described in
Section~5) produces a sequence of $15$ integers that is simultaneously
non-averaging and 2-non-averaging over $\Z_q$. However, the theorem
applied with $\ell=2$, $c_1 = 1$ and $c_2 = 2$ gives a lower bound
of $8$ for the cardinality of such a sequence.

A more general lower bound can be derived by extending a result of
Behrend \cite{Beh46} derived originally for non-averaging sequences. Consider
the following system, $\Omega$, of $\ell$ equations in the variables
$\sfu_1,\sfu_2,\ldots,\sfu_m,\sfv$:
\begin{equation}\label{Omega}
\Omega\  :\  \begin{array}{c}
  \sum_{j = 1}^m\, c_{1,j} \, \sfu_j = b_1 \, \sfv \\
  \ldots \\
  \sum_{j = 1}^m \, c_{\ell,j} \, \sfu_j = b_{\ell} \, \sfv, \\
\end{array}
\end{equation}
where the coefficients $c_{i,j}, b_i$ are non-negative integers
such that for each $i \in [1,\ell]$, at least two of the
$c_{i,j}$'s are nonzero, and $\sum_{j = 1}^m c_{i,j} = b_i > 0.$

\begin{theorem}
Given a system, $\Omega$, as in (\ref{Omega}),
let $D=\max_{1 \leq i \leq \ell}\,b_i$. Then, for $q > D^2$,
$$
s(q;\O) \geq
\gamma_1 \, q \, e^{-\gamma_2 \sqrt{\log q} - \frac{1}{2} \log\log q} \ (1+o(1))
$$
where $\log$ denotes the natural logarithm,
$\gamma_1 = D^2 \sqrt{\frac{1}{2}\log D}$, $\gamma_2 = 2\sqrt{2\log D}$,
and $o(1)$ denotes a correction factor that vanishes as $q \rightarrow \infty$.
\label{Omega_theorem}
\end{theorem}

We postpone the proof of the theorem to the Appendix.
The above result can be compared directly to the result of
Theorem~\ref{c-non-av_theorem} since the system of equations
$\sfx + c_i\sfy = (c_i+1)\sfz$, $i = 1,2,\ldots,\ell$,
is of the form given in \eqref{Omega}. Therefore, the result of
Theorem~\ref{Omega_theorem} applies to this system of equations $\O$,
with $D = 1 + \max_i c_i$. It is easily seen that
by the bound of Theorem~\ref{Omega_theorem},
$$
\lim_{q \rightarrow \infty} \frac{s(q;\O)}{q^{1-\epsilon}} = \infty
$$
for any $\epsilon > 0$. Since $\epsilon$ can be chosen to be
arbitrarily small, this is much stronger, asymptotically in $q$,
than the result of Theorem~\ref{c-non-av_theorem}, which only shows that
$s(q;\O) \geq C\,q^{1/3}$ for some constant $C > 0$ independent of $q$.
However, for small values of $q$, particularly
for the values of the modulus $q$ typically used in practical array code
design, the bound of Theorem~\ref{c-non-av_theorem} is better
than that of Theorem~\ref{Omega_theorem}. For instance, when applied
to the system, $\O$, consisting of the pair of equations
$\sfx + \sfy = 2 \sfz$ and $\sfx + 2 \sfy = 3 \sfz$,
the bound of Theorem~\ref{Omega_theorem}, for
$q = 241$, evaluates to $0.66$, which just shows that $s(q;\O) \geq 1$.
As stated earlier, the bound of Theorem~\ref{c-non-av_theorem} yields
$s(q;\O) \geq 8$ in this case.

To conclude this section, we remark that while the problem of
precisely estimating the growth rate of $s(q;\O)$ with $q$ is one of
considerable interest and value, finding provably good estimates is a
notoriously difficult problem. For example, the current best
lower bound for the growth rate of the cardinality of
non-averaging sequences is that due to Behrend
(Theorem~\ref{Omega_theorem} for the special case of $\O$ consisting
of the single equation $\sfx + \sfy = 2 \sfz$),
but it is still not known whether this is the best possible such bound.


\section{Construction Methods}

The simplest and computationally least expensive methods for
generating integer sequences satisfying a given set of constraints
are greedy search strategies and variations thereof. A typical
greedy search algorithm starts with an initial \emph{seed} sequence
that trivially satisfies the given constraints, and progressively
extends the sequence by adding new terms that continue to maintain
the constraints.

As an example, to construct a non-negative integer sequence that
contains no solutions to any equation within a system, $\Omega$, of
cycle-governing equations of the form \eqref{part_reg_eq}, we start
with a seed sequence of $m-1$ non-negative integers, $n_1 < n_2 <
\ldots < n_{m-1}$, where $m$ is the least number of variables among
any of the equations in $\Omega$. For each $j \geq m$, we take $n_j$
to be the least integer greater than $n_{j-1}$ such that
$\{n_1,n_2,\ldots,n_j\}$ contains no solutions to any equation in
$\Omega$. The rate of growth of elements in a sequence generated by
such a greedy search procedure is influenced by the choice of the
seed sequence \cite{OS78}. The search needs to be performed only once to
generate a sequence of integers avoiding solutions to any equation in
$\O$, and it is easily seen that the algorithm
has complexity $O(L\cdot q^{M+1})$,
where $L$ denotes the number of equations in $\O$,
$M$ is the maximum number of variables
among all these equations, and $q$ is the prime modulus.
Tables~\ref{PACr3_table}--\ref{IACr4_table} list the output of the
greedy search procedure, initialized by different seed sequences,
for finding sequences that avoid solutions to various
cycle-governing equations in PAC's and IAC's. The first two terms of
each sequence listed in the tables form the seed sequence for the
greedy search algorithm.

\begin{table*}[tbp]
\caption{Cycle-governing equations over $\Z_q$ for PAC's with modulus $q$
and column-weight $r=3$, and greedy sequences avoiding solutions over
$\Z_{1213}$ to them.}
\vspace{0.2in}
\centering
\begin{tabular}{||c|c||}
\hline
Six-cycle equation & Greedy sequences avoiding the six-cycle equation \\
\hline
$ 2i-j-k=0 $
&
$
\begin{array}{l}
    0,1,3,4,9,10,12,13,27,28,30,38,\ldots \\
    0,2,3,5,9,11,12,14,27,29,30,39,\ldots \\
    0,3,4,7,9,12,13,16,27,30,35,36,\ldots
\end{array}
$ \\
\hline
Eight-cycle equations & Greedy sequences avoiding all six- and
eight-cycle equations \\
\hline
$
\begin{array}{rcl}
    2i+j-k-2l&=&0 \\
    i+j-k-l&=&0 \\
    3i-j-2k&=&0 \\
    2i-j-k&=&0
\end{array}
$
&
$
\begin{array}{l}
    0,1,4,11,27,39,48,84,134,163,223,284,333,\ldots \\
    0,2,5,13,20,37,58,91,135,160,220,292,354,\ldots \\
    0,3,4,13,25,32,65,92,139,174,225,318,341,\ldots
\end{array}
$ \\
\hline
\end{tabular}
\label{PACr3_table}
\end{table*}

\begin{table*}[tbp]
\begin{center}
\caption{Cycle-governing equations over $\Z_q$ for PAC's with modulus $q$
and column-weight $r=4$, and greedy sequences avoiding solutions over
$\Z_{911}$ to them.}
\vspace{0.2in}
\begin{tabular}{||c|c|c||}
\hline
Six-cycle equations & Greedy sequences avoiding all six-cycle equations \\
\hline
$
\begin{array}{rcl}
     2i-j-k&=&0 \\
     3i-j-2k&=&0
\end{array}
$
&
$
\begin{array}{l}
    0,1,4,5,11,19,20,\ldots \\
    0,2,5,7,13,18,20,\ldots \\
    0,3,4,7,16,17,20,\ldots
\end{array}
$ \\
\hline
Eight-cycle equations & Greedy sequences avoiding all six- and
eight-cycle equations \\
\hline
$
\begin{array}{rcl}
    3i-j-k-l&=&0 \\
    3i-2j-2k+l&=&0 \\
    2i-2j-k+l&=&0 \\
    3i-3j+k-l&=&0 \\
    3i-3j+2k-2l&=&0 \\
    i+j-k-l&=&0 \\
    2i-j-k&=&0\\
    4i-3j-k&=&0\\
    3i-2j-k&=&0\\
    5i-3j-2k&=&0
\end{array}
$
&
$
\begin{array}{l}
    0,1,5,18,25,62,95,148,207,\ldots \\
    0,2,7,20,45,68,123,160,216,\ldots \\
    0,3,7,22,39,68,123,154,244,\ldots
\end{array}
$ \\
\hline
\end{tabular}
\label{PACr4_table}
\end{center}
\end{table*}

\begin{table*}[tbp]
\begin{center}
\caption{Cycle-governing equations over $\Z_q$ for IAC's with modulus $q$,
column-weight $r=3$ and block-row labels $\{0,1,3\}$,
and greedy sequences avoiding solutions over $\Z_{1213}$ to them.}
\vspace{0.3in}
\centering
\begin{tabular}{||c|c||}
  \hline
Six-cycle equation & Greedy sequences avoiding the six-cycle equation \\
  \hline
$
\begin{array}{c}
3i-2j-k=0
\end{array}
$
&
$\begin{array}{l}
0,1,2,5,8,9,10,16,18,21,33,35,37,40,\ldots \\
0,2,4,7,9,11,14,16,18,31,35,39,45,\ldots \\
0,3,4,5,8,11,13,19,20,21,32,36,40,\ldots
\end{array}
$ \\
\hline
Eight-cycle equations & Greedy sequences avoiding all six- and
eight-cycle equations \\
\hline
$
\begin{array}{rcl}
    3i-3j-k+l&=&0\\
    3i-3j-2k+2l&=&0\\
      i+j-k-l&=&0\\
    2i+j-k-2l&=&0\\
    4i-3j-k&=&0\\
    2i-j-k&=&0 \\
    5i-3j-2k&=&0
\end{array}
$
&
$
\begin{array}{l}
    0,1,5,14,25,57,88,122,198,257,280,\ldots \\
    0,2,7,18,37,65,99,151,220,233,545,\ldots \\
    0,3,7,18,31,50,105,145,186,230,289,\ldots
\end{array}
$ \\
\hline

Ten-cycle equations & Greedy sequences avoiding six-, eight- and
ten-cycle equations \\
\hline
$
\begin{array}{rcl}
    3i-j+k-l-2m&=&0 \\
    3i-j-2k+2l-2m&=&0 \\
    3i+j+2k-3l-3m&=&0 \\
    3i-j-k-l&=&0 \\
    3i-3j-k+l&=&0 \\
    3i-2j+k-2l&=&0 \\
    i-4j+k+2l&=&0 \\
    3i-j-5k+3l&=&0 \\
    3i-j-4k+2l&=&0 \\
    i-2j+2k-l&=&0 \\
    3i-2j+2k-3l&=&0 \\
    6i-j-2k-3l&=&0 \\
    5i-j-2k-2l&=&0 \\
    4i-3j-3k+2l&=&0 \\
    3i-2j-k&=&0 \\
    i-4j+3k&=&0 \\
    3i+2j-5k&=&0\\
    2i-j-k&=&0 \\
    6i-j-5k&=&0\\
    5i-j-4k&=&0
\end{array}
$
&
$
\begin{array} {l}
0,1,7,29,96,148,324 \\
0,2,7,29,70,178,733 \\
0,3,7,26,54,146,237
\end{array}
$ \\
\hline
\end{tabular}
\label{IACr3_table}
\end{center}
\end{table*}

\begin{table*}[tbp]
\begin{center}
\caption{Cycle-governing equations over $\Z_q$ for IAC's with modulus $q$,
column-weight $r=4$ and block-row labels $\{0,1,3,7\}$,
and greedy sequences avoiding solutions over $\Z_{911}$ to them.}
\vspace{0.3in}
\begin{tabular}{||c|c||}
\hline
Equations (Six-cycles) & Greedy sequences avoiding all six-cycle equations \\
\hline
$
\begin{array}{rcl}
     3i-j-2k&=&0\\
     7i-j-6k&=&0\\
     7i-3j-4k&=&0
\end{array}
$
&
$
\begin{array}{l}
    0,1,2,5,10,12,19,25,27,41,42,46,50,60,\ldots \\
    0,2,4,9,10,17,20,34,36,45,55,61,71,77,\ldots \\
    0,3,4,5,8,13,20,27,37,46,47,48,51,66,\ldots
\end{array}
$ \\
\hline
Equations (Eight-cycles) & Greedy sequences avoiding all six- and
eight-cycle equations \\
\hline
$
\begin{array}{rcl}
    7i-4j-2k-l&=&0\\
    7i-6j-3k+2l&=&0\\
    7i-7j-k+l&=&0\\
    7i-7j+3k-3l&=&0\\
    7i-7j+6k-6l&=&0\\
    7i-7j+4k-4l&=&0\\
    6i-6j-k+l&=&0\\
    6i-4j-3k+l&=&0\\
    4i-4j-3k+3l&=&0\\
    3i-3j-2k+2l&=&0\\
    3i-3j-k+l&=&0\\
    2i-2j-k+l&=&0\\
    i+j-k-l&=&0\\
    9i-7j-2k&=&0\\
    7i-5j-2k&=&0\\
    5i-4j-k&=&0\\
    4i-3j-k&=&0\\
    3i-2j-k&=&0\\
    2i-j-k&=&0\\
    5i-3j-2k&=&0\\
    8i-7j-k&=&0\\
    6i-5j-k&=&0\\
    13i-7j-6k&=&0\\
    10i-7j-3k&=&0\\
    11i-7j-4k&=&0
\end{array}
$
&
$
\begin{array}{l}
    0,1,9,20,46,51,280 \\
    0,2,11,19,42,83,118 \\
    0,3,8,25,45,72,142
\end{array}
$ \\
\hline
\end{tabular}
\label{IACr4_table}
\end{center}
\end{table*}

There is an alternative procedure that often generates sequences with
more terms than a simple greedy search routine. The
idea is to start with some construction of a dense sequence
avoiding solutions to some subset of the cycle-governing
equations in the set $\Omega$,
and then to sequentially expurgate elements of that
sequence that violate any of the remaining constraints. After the expurgation
procedure is completed, additional elements may be added to the sequence
as long as they jointly avoid solutions to all cycle-governing
equations in $\Omega$.

A good sequence with which to start this alternative procedure
can be constructed according to a method outlined by Bosznay
\cite{Bos89}. The construction proceeds through the following steps.
First, a prime $q$ is chosen, and along with it the smallest
integer $t$ such that $q \leq t^4$. Let
$$
  n_j=j\,t^3+\frac{j(j+1)}{2}, \; \;\;\, j=1,2,\ldots,t-1,
$$
and let $S' = \{n_1,n_2,\ldots,n_{t-1}\} \cap [0,q-1]$. It
can be shown that the sequence $S'$ does not contain proper solutions
over $\Z_q$ to any equation of the form
$$
\sum_{i=1}^m \, c_i \, \sfu_i = b \, \sfv
$$
where $c_1,c_2,\ldots,c_m,b$ are positive integers such that
$\sum_{i=1}^m \, c_i = b$.
Next, one uses a simple greedy algorithm to find the largest
subset $S \subset S'$ that does not contain proper solutions to
cycle-governing equations in $\O$ that are not of the above form.
The last step in the procedure is
to check whether there exist integers in $[0,q-1]$ that can be added to $S$
without creating a proper solution within $S$ to some cycle-governing equation.
If such integers exist, they are sequentially added to the set $S$.

As illustrative examples, we list three sequences constructed using
the adaptation of Bosznay's method described above. The sequence
$1,4,8,23,40,126,253,352,381,495$ constructed by this method does
not contain solutions to any of the equations listed in
Table~\ref{PACr4_table} that govern cycles of length six and eight
in a PAC with modulus $q=911$ and column-weight $r=4$. In
comparison, the greedy algorithm initialized by the seed sequence
$0,1$ produces $0,1,5,18,25,62,95,148,207$. The sequences
$6,8,165,217,435,654,1095$ and $0,1,7,29,64,111,753$, generated by
the modified Bosznay construction and the greedy algorithm with seed
sequence $0,1$, respectively, avoid solutions to any of the
equations listed in Table~\ref{IACr3_table}. Finally, in the case of
the equations in Table~\ref{IACr4_table}, the sequences produced by
the two methods are $2,4,28,217,255,435,654$ and $0,1,9,20,46,51$.
Observe that the sequences produced by the modified Bosznay
construction contain terms that are larger in general than the terms
in the corresponding greedy sequences where almost all elements are
much smaller than the prime $q$.

\section{Simulation Results}

In this section, we present the bit-error-rate (BER)
curves over an AWGN channel for various (shortened) PAC's and IAC's,
and also provide comparisons with other codes of similar rates and lengths
from the existing literature. All array codes considered in this section
were iteratively decoded using a sum-product/belief-propagation (BP) decoder.

\begin{figure}
\centering
\epsfig{file=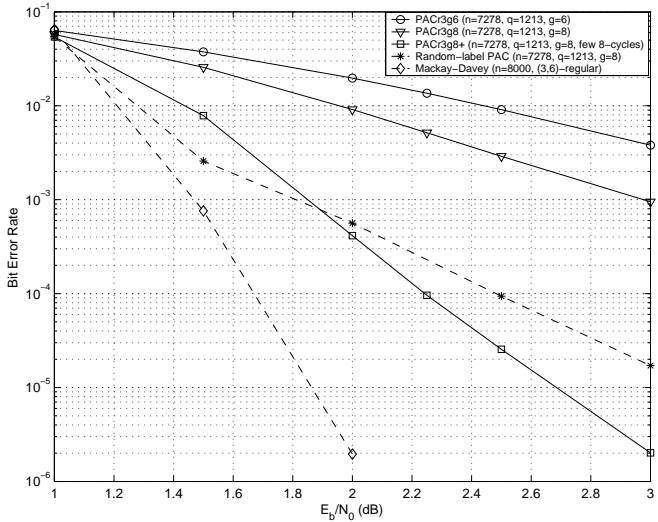, width=8.8cm}
   \caption{\label{performance1}
        BER versus $E_b/N_0$ (dB) for designed rate-1/2 PAC's with $r=3$.}
\end{figure}

\begin{figure}[tp]
\centering
\epsfig{file=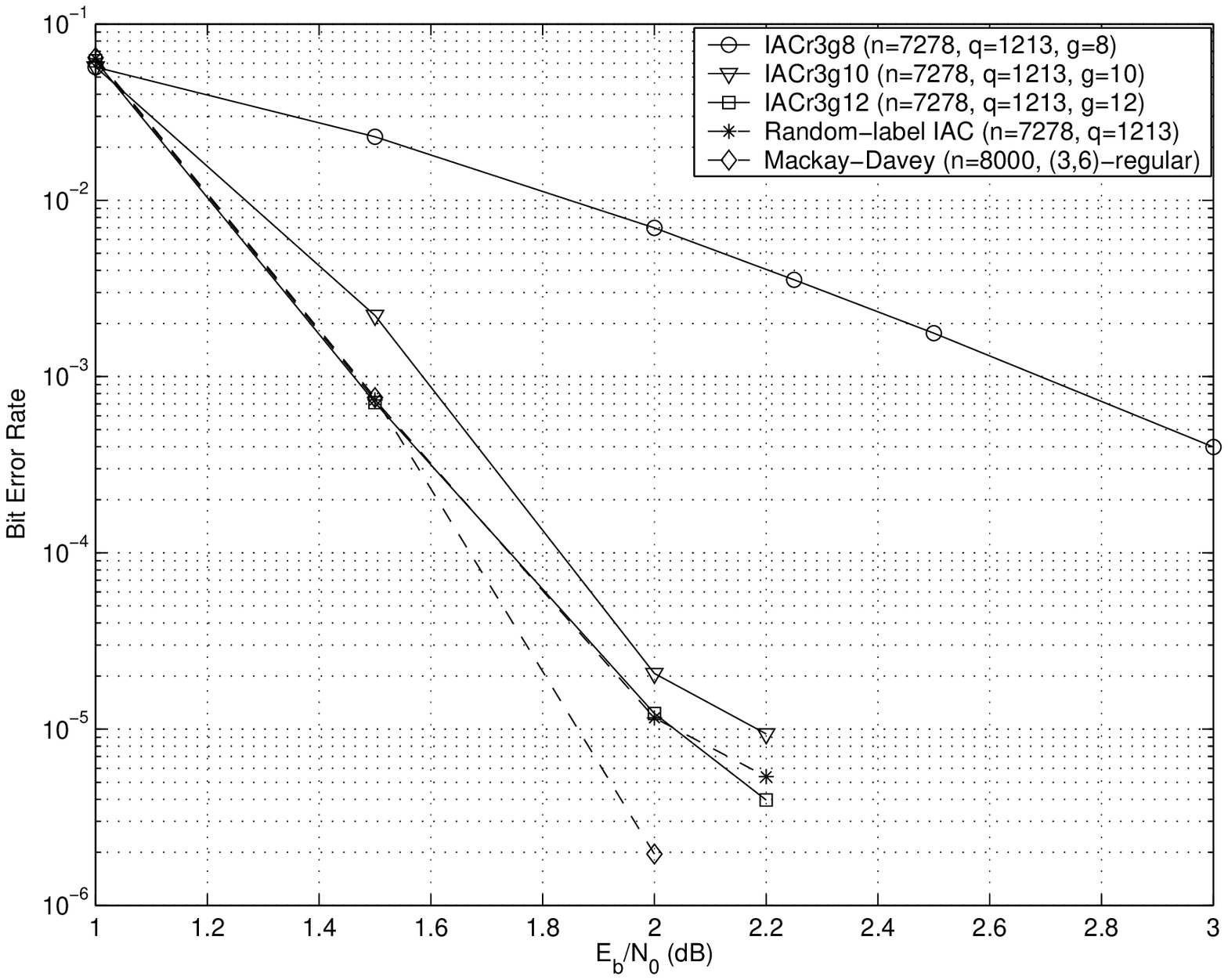, width=8.8cm}
   \caption{\label{performance2}
        BER versus $E_b/N_0$ (dB) for designed rate-1/2 IAC's with $r=3$.}
\end{figure}

Figures~\ref{performance1} and \ref{performance2} show the
performance curves, after a maximum of 30 rounds of iterative
decoding, for array codes of column-weight 3 and row-weight 6; thus
all these codes have designed rate 1/2. The prime modulus used for
the construction of these codes is $q=1213$, which yields codes with
length 7278. The sets of block-column labels used in the codes
PACr3g6, PACr3g8 and PACr3g8+ in Figure~\ref{performance1} are
$\{0,1,2,3,4,5\}$, $\{0,1,3,4,9,10\}$ and $\{0,1,4,11,27,39\}$,
which correspond to a PAC of girth six, a shortened PAC of girth
eight, and a shortened PAC of girth eight but without eight-cycles
governed by the equations in Table~\ref{PACr3_table}, respectively.
The codes IACr3g8, IACr3g10 and IACr3g12, whose performance is
plotted in Figure~\ref{performance2}, are of girth eight, ten and
twelve, respectively. The respective sets of block-column labels are
$\{0,1,2,5,7,8\}$, $\{0,1,5,14,25,57\}$, and $\{0,1,7,29,64,111\}$.
All the IAC's in the figure have block-row labels $\{0,1,3\}$.

\begin{figure}[tp]
\centering
\epsfig{file=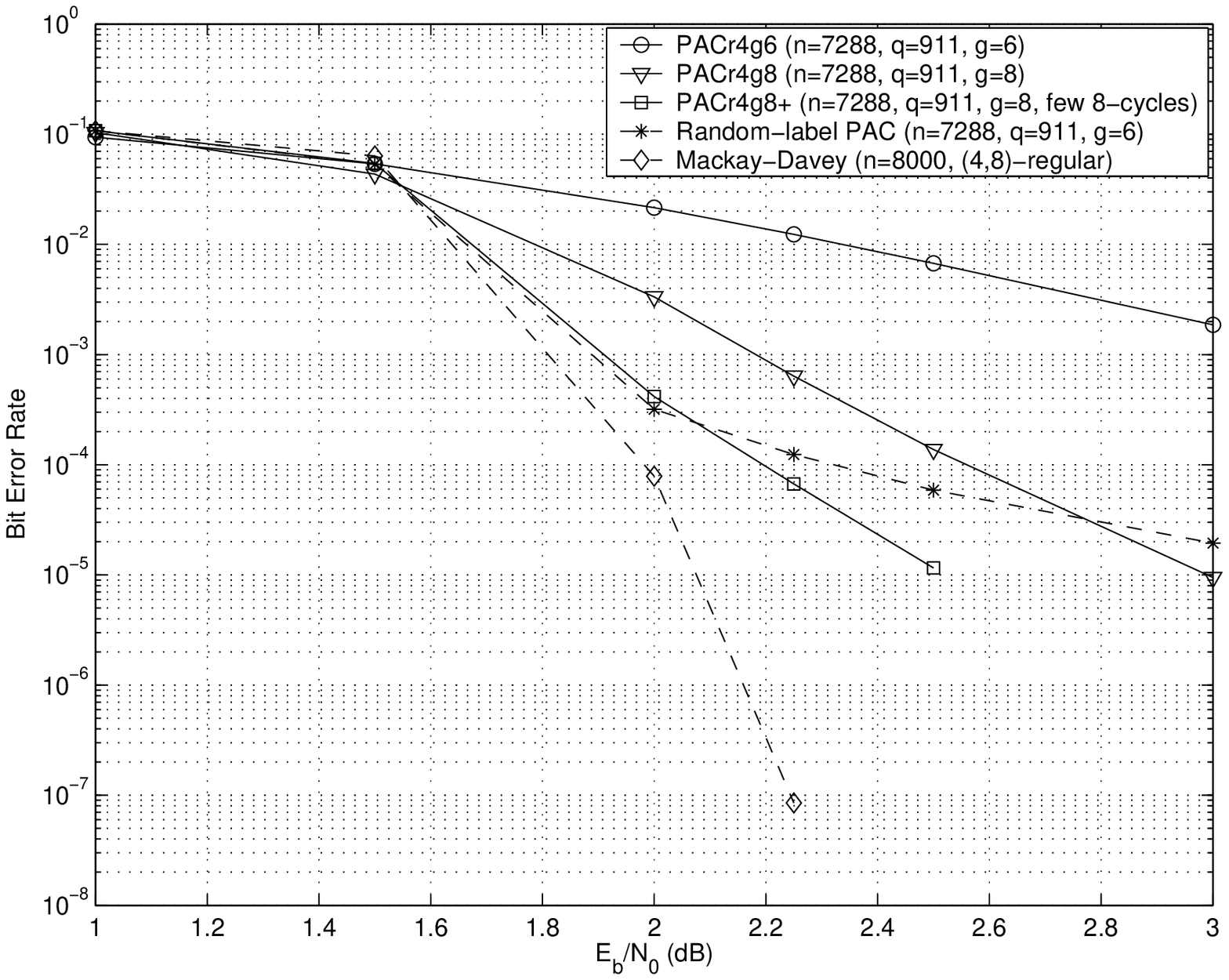, width=8.8cm}
\caption{\label{performance3}
        BER versus $E_b/N_0$ (dB) for designed rate-1/2 PAC's with $r=4$.}
\end{figure}

\begin{figure}[tp]
\centering
\epsfig{file=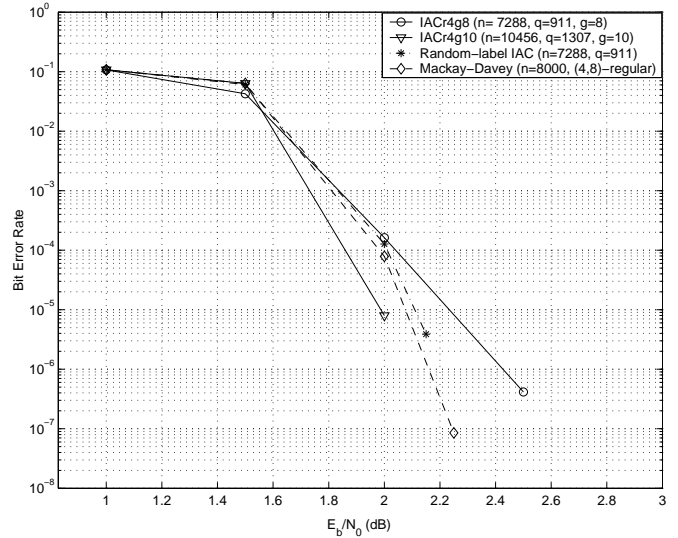, width=8.8cm}
\caption{\label{performance4}
        BER versus $E_b/N_0$ (dB) for designed rate-1/2 IAC's with $r=4$.}
\end{figure}

Figures~\ref{performance3} and \ref{performance4} show the results,
after a maximum of 30 decoding iterations, for codes with designed
rate 1/2 and column-weight $r=4$. The array codes in
Figure~\ref{performance3} are shortened PAC's with modulus $q=911$
and length $7288$. The sequences used for the block-column labels in
the codes PACr4g6, PACr4g8 and PACr4g8+ are $\{0,1,2,3,4,5,6,7\}$,
$\{0,3,4,7,16,17,20,22\}$ and $\{0,1,5,18,25,62,95,148\}$,
respectively. The codes PACr4g6 and PACr4g8 are of girth six and
eight, respectively, while PACr4g8+ is a code of girth eight with no
eight-cycles governed by the equations in Table~\ref{PACr4_table}.
The codes IACr4g8 and IACr4g10 in Figure~\ref{performance4} are
IAC's of girth eight and ten, respectively, that use the set of
block-row labels $\{0,1,3,7\}$, but differ in the modulus and
block-column labels used. The code of girth eight has modulus
$q=911$, hence length 7288, and block-column labels
$\{0,1,2,5,9,10,18,42\}$. The girth-ten code, on the other hand,
uses the modulus $q=1307$, so that it has length 10456, and
block-column labels $\{317,344,689,1035,
1178,1251,1297,1303\}$ 
The reason for not choosing $q$ to be 911
in the girth-ten code is that none of the construction methods discussed
in Section~5 produces a sequence of length eight without solutions over
$\Z_{911}$ to any of the equations listed in Table~\ref{IACr4_table}.
The smallest choice for the prime $q$ which does produce a
sequence of eight block-column labels satisfying the eight-cycle constraints
turns out to be 1307.

For comparison purposes, each of
Figures~\ref{performance1}--\ref{performance4} also contains the BER
curves for two other codes: a designed rate-1/2, regular LDPC code
of length 8000 with a random-like structure, as constructed by
MacKay and Davey in \cite{Mac03}, and a ``random-label'' array code
in which the block-row and block-column labels are randomly chosen.
The MacKay-Davey code in Figures~\ref{performance1} and
\ref{performance2} is a $(3,6)$-regular code, while that in
Figures~\ref{performance2} and \ref{performance4} is a
$(4,8)$-regular code. The random-label code in
Figure~\ref{performance1} is a PAC with $q = 1213$, $r=3$ and set of
block-column labels $\{24, 460, 610, 826, 1009, 1012\}$. Among the
equations in Table~\ref{PACr3_table}, this label set contains
solutions over $\Z_{1213}$ to only one equation, namely,
$3i-2j-k=0$; the solution is $(i,j,k) = (826,1009,460)$. Thus, this
PAC contains no six-cycles and relatively few eight-cycles. The
random-label code in Figure~\ref{performance2} is an IAC with the
same choices of $q$, $r$ and block-column labels as in the
random-label PAC above, but the block-row label set for the code is
$\{3,4,7\}$. The random-label PAC in Figure~\ref{performance3} and
the random-label IAC in Figure~\ref{performance4} have $q = 911$,
$r=4$ and set of block-column labels $\{17, 210, 415, 442, 552, 694,
811, 865\}$; the IAC has block-row labels $\{2,5,7,8\}$. The set of
block-column labels for the random-label PAC in
Figure~\ref{performance3} supports proper solutions over $\Z_{911}$
to several of the equations in Table~\ref{PACr4_table}. These
equations and solutions are tabulated in Table~\ref{eq_sol_table}.
It is clear that this array code contains many six-cycles and
eight-cycles.

\begin{table}[ht]
\caption{Solutions over $\Z_{911}$ supported within the set
$\{17, 210, 415, 442, 552, 694, 811, 865\}$ to the cycle-governing
equations in Table~\ref{PACr4_table}.}
\label{eq_sol_table}
\begin{center}
\begin{tabular}{||r|c||}
\hline
Equation \ \ \ \ \ \ \ \ & Solutions $(i,j,k)$ or $(i,j,k,l)$ \\
\hline
$2i-j-k = 0$ & $(811,17,694)$, $(811,694,17)$ \\[4pt]
$2i-2j-k+l=0$ & $(415,442,811,865)$, $(442,415,865,811)$, \\
& $(694,865,210,552)$, $(865,694,552,210)$ \\[4pt]
$3i-2j-2k+l=0$ & $(865,694,811,415)$, $(865,811,694,415)$ \\[4pt]
$3i-3j+2k-2l = 0$ & $(210,552,17,415)$, $(552,210,415,17)$ \\[4pt]
$3i-3j+k-l = 0$ & $(694,865,17,415)$, $(865,694,415,17)$ \\[4pt]
\hline
\end{tabular}
\end{center}
\end{table}

From the simulation results presented in
Figures~\ref{performance1}--\ref{performance4}, we can clearly observe
the sharp improvement in performance that can be achieved by increasing the
girth of an array code, or even by partially eliminating cycles of
a certain fixed length. As girth increases, the BER curves of
array codes approach that of a random-like LDPC code of similar
length and degree distribution. This provides concrete evidence in
support of the widely-held belief that the girth of a code is
an important factor in determining its performance. This also appears
to be borne out by the performance of the random-label PAC's in
Figures~\ref{performance1} and \ref{performance3}.
As can be seen from these figures, the degradation in
performance (in comparison with the random-like MacKay-Davey codes)
of the random-label PAC of column-weight three is
significantly smaller than that of column-weight four. Recall
that the random-label PAC of column-weight three contains few short cycles,
while the code of column-weight four contains many six-cycles and eight-cycles.

The best performance among the array codes we considered for our
simulations was achieved by IACr4g10, for which there was no observed
error for $50$ million simulated blocks and 30 iterations of
message-passing, implying that at an SNR of $2.5$dB, the BER
achieved by the code is less than $10^{-9}$. As can be seen from
Figure~\ref{performance4}, this code performs better than the
random-like MacKay-Davey code of comparable parameters.

It is worth pointing out that the PAC's of column-weight four and
the IAC of girth eight and column-weight four significantly
outperform their counterparts with column-weight three. This is the
reverse of the trend observed among LDPC codes with random-like
structure, as it is known that among such codes, $(3,6)$-regular
codes have the best threshold properties at rates below $0.9$, as
can be clearly seen from the performance plots of the random-like
MacKay-Davey codes in Figures~\ref{performance1} and
\ref{performance3}. We conjecture that the observed results are a
consequence of the fact that the array codes of designed rate 1/2,
length around $8000$ and column-weight four have minimum distance
significantly larger than their column-weight three counterparts, or
that they have relatively few cycles of length equal to or exceeding
the girth, and probably have almost optimal structure, (\emph{i.e.}\
they are comparable to random-like codes). At the same time, array
codes with designed rate 1/2 and column-weight three show a
significant gap away from the optimal performance, since for such a
degree distribution it is very likely that optimal LDPC codes can
have girth much larger than twelve, and larger minimum distance than
the upper bound listed in Table~\ref{dmin_table}.

Finally, we provide some data comparing the performance of shortened
array codes with that of some of the structured LDPC codes studied
in the existing literature. We start with the class of LDPC codes
derived in \cite{KLF01} from projective and Euclidean geometries
over finite fields. Most of these codes have much higher rates than
the shortened array codes with comparable codelengths. Shortened
array codes of a certain codelength tend not to achieve rates as
high as those achieved by codes of the same length derived from
projective and Euclidean geometries due to the relatively small
density of integer sequences avoiding solutions to cycle-governing
equations. So, to make a fair comparison, we consider, as an
example, the code of length 8190 and dimension 4095 obtained by
``extending'' the $(4095,3367)$ Type-I 2-dimensional Euclidean
geometry code via the column-splitting procedure described in
\cite[Section~VI]{KLF01}. This code has rate 1/2, and so can be
compared with the designed rate-1/2 shortened array codes of similar
lengths. As reported in \cite[Table~III]{KLF01}, a shortened
projective geometry code with parameters $(8190,4095)$ achieves a
BER of $10^{-4}$ at an SNR of 6 dB, which is $5.82$ dB away from the
Shannon limit of 0.18 dB. On the other hand, the length-7288,
designed rate-1/2 code IACr4g8 in Figure~\ref{performance4} achieves
the same BER at an SNR of slightly less than 2dB, which is
considerably closer to the Shannon limit.

\begin{figure}[tp]
\centering \epsfig{file=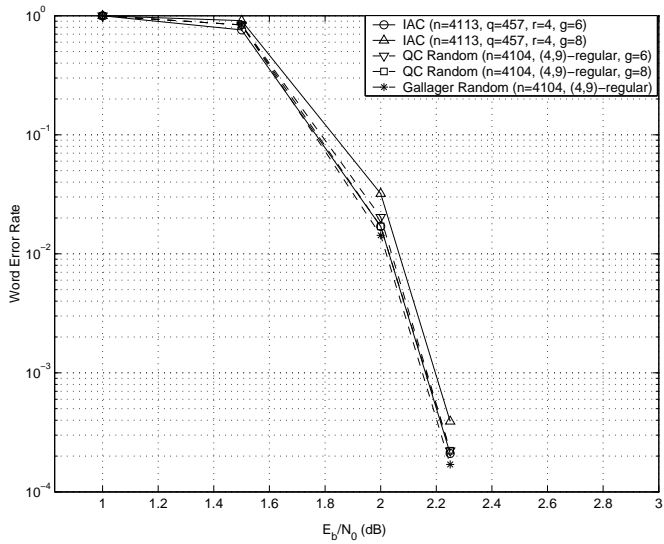, width=8.8cm}
\caption{Comparison of array codes with some quasi-cyclic codes from
\cite{Fos04}. All codes in the figure have lengths around 4100, and are
(4,9)-regular.}
\label{performance5}
\end{figure}

\begin{figure}[tp]
\centering \epsfig{file=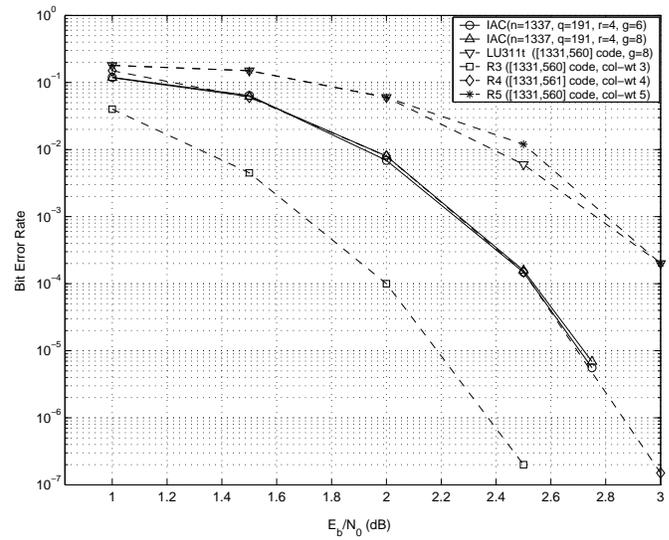, width=8.8cm}
\caption{Comparison of array codes with some codes from
\cite{KPP04}. All codes being compared have lengths around 1330 and
designed rates 0.42--0.43.} \label{performance6}
\end{figure}

Figures~\ref{performance5} and \ref{performance6} provide a
comparison of the performance of array codes with the codes studied
in \cite{Fos04} and \cite{KPP04}. The first of these figures
compares the performance of a pair of IAC's with a pair of random
quasi-cyclic codes and a random Gallager code from
\cite[Figure~2]{Fos04}. All the codes in the figure have lengths
around 4100, and are (4,9)-regular, hence have designed rate 5/9.
Both the IAC's plotted have length 4113, modulus $q=457$,
column-weight $r=4$, and block-row labels $\{0,1,3,7\}$. The set of
block-column labels used in the two IAC's are
$\{0,1,9,10,22,31,32,172,194\}$ and $\{0,1,9,10,24,43,88,90,326\}$,
respectively. The first sequence avoids solutions over $\Z_{457}$ to
all the equations governing six-cycles listed in
Table~\ref{IACr4_table}, except for $7i-3j-4k = 0$ which has three
solutions --- $(9, 172, 1)$, $(10, 22, 1)$ and $(22, 10, 31)$ ---
within the sequence. The code corresponding to this sequence thus
has girth six. Of the 25 equations governing eight-cycles listed in
Table~\ref{IACr4_table}, the sequence contains solutions over
$\Z_{457}$ to exactly 11 equations, the number of solutions to these
equations being 50 in all. The sequence
$\{0,1,9,10,24,43,88,90,326\}$ contains no solutions over $\Z_{457}$
to any of the six-cycle equations in Table~\ref{IACr4_table}, but
contains a total of 68 solutions to 14 of the eight-cycle equations.
Thus the IAC with this set of block-column labels has girth eight,
but has considerably more cycles of length up to eight than the IAC
with the first set of labels, and so performs somewhat worse (see
Figure~\ref{performance5}). Overall, the performance (in terms of
word error rate) of all the codes in Figure~\ref{performance5} codes
is quite similar, but it should be noted that the plotted
performance of the two IAC's was obtained after a maximum of 50
rounds of BP decoding, while the codes from \cite{Fos04} were
allowed a maximum of 200 rounds of BP decoding.

In Figure~\ref{performance6}, a pair of IAC's is compared with several
codes of similar lengths and rates taken from \cite{KPP04}.
The IAC's in the plot all have length 1337, modulus $q=191$,
column-weight $r=4$ and block-row labels $\{0,1,3,7\}$. They differ
in the sequence of block-column labels used: one uses the sequence
$\{0,1,9,10,22,31,126\}$, while the other uses $\{0,1,5,6,25,46,151\}$.
The former sequence contains solutions over $\Z_{191}$ to exactly
one six-cycle equation from Table~\ref{IACr4_table} --- the
solutions are $(0, 126, 1)$, $(10, 22, 1)$, $(22, 10, 31)$ and $(126, 10, 22)$
--- and a total of 44 solutions to 15 eight-cycle equations.
Thus, this sequence yields a girth-six code. The sequence
$\{0,1,5,6,25,46,151\}$ contains solutions over $\Z_{191}$ to none
of the six-cycle equations in Table~\ref{IACr4_table}, and
altogether 50 solutions to 15 eight-cycle equations. Thus, despite
the fact that the IAC with this sequence of block-column labels has
girth eight, its performance is almost identical to that of the
girth-six IAC, which can be explained by the fact that
the two codes have a similar cycle distribution.
The code $LU311t$ is a structured LDPC code based on a construction
of a family of regular bipartite graphs by Lazebnik and Ustimenko
\cite{LU97}. The parity-check matrix of the code is a
$1331 \times 1331$ square matrix with row-weight and column-weight 11.
The code has girth eight, dimension 560 and minimum distance at least 22.
\cite{KPP04}. The codes $R3$, $R4$ and $R5$ are irregular
random-like LDPC codes with parity-check matrices of column-weight
3, 4 and 5 respectively. The performance plots of the codes $LU311t$,
$R3$, $R4$ and $R5$ have been obtained from \cite[Figure~5]{KPP04}, where
it is stated that a maximum of 500 iterations of BP decoding was allowed
for each of these codes.
The performance of the IAC's in the Figure was obtained after a maximum
of 50 rounds of BP decoding. As can be seen from the figure, the
two IAC's match the performance of the random-like
column-weight-four code $R4$, and easily outperform the code $LU311t$.


\section{Conclusion}

In summary, in this paper, we considered the problem of constructing new LDPC
codes with large girth based on the array code construction of \cite{Fan00}.
Our contributions were threefold. Firstly, we provided a simple method
for relating cycles in the Tanner graph of such codes to homogeneous
linear ``cycle-governing'' equations with integer coefficients.
This yields an approach for constructing codes
with a desired cycle distribution, based on the existence of
integer sequences that avoid solutions to the cycle-governing equations.
Secondly, we provide some bounds on the cardinality of integer
sequences avoiding solutions to such equations, which give useful
esimates of the rate penalty incurred in shortening an array code
to eliminate cycles. Finally, we showed through extensive simulations
the influence of various kinds of short cycles on the performance of
LDPC codes under iterative decoding.

\section*{Appendix}
\renewcommand{\thesection}{A}

We provide proofs of Lemma~\ref{roth_sidon_lemma}(b) and
Theorems~\ref{c-non-av_theorem} and \ref{Omega_theorem} in this Appendix. \\

{\em Proof of Lemma~\ref{roth_sidon_lemma}(b)\/}: Let $S$ be a Sidon sequence
over $\Z_N$, and let $\mathcal{P}$ be the set $\{(a,b): a,b \in S,
a \neq b\}$. From the definition of a Sidon sequence and the fact
that $N$ is odd, it follows that the mapping $f: \mathcal{P}
\rightarrow [1,N-1]$ defined by $f(a,b) = a-b \mod N$ is
injective. Therefore, $N-1 \geq |\mathcal{P}| = |S| (|S|-1)$.
Solving the associated quadratic equation,
we obtain $|S| \leq \sqrt{N-3/4} + 1/2$. \endproof  \mbox{ }

For the proof of Theorem~\ref{c-non-av_theorem}, we recall some definitions
from graph theory. A \emph{hypergraph}, $H =(V,E)$, is an ordered pair
of two finite sets: the set of vertices $V$, and the set of edges
$E$, which are arbitrary non-empty subsets of $V$. A hypergraph is
called {\em $h$-uniform} if all its edges have the same
cardinality $h$, and is called {\em $s$-regular} if all its
vertices belong to the same number, $s$, of edges.
A set of vertices of a hypergraph, $H$, which does not (completely)
contain any edge of $H$ is called an {\em independent set}.
The maximum cardinality of an independent set of $H$
is called the {\em independence number} of $H$, and is denoted
by $\alpha(H)$. \\

\emph{Proof of Theorem~\ref{c-non-av_theorem}\/}:
Let $\O$ be as in the statement of the theorem.
Define a hypergraph $H(q;\O)$ with vertex set $[0,q-1]$,
and a set of edges that consists of all triples of the form
$$
\{x, \, x+t \!\!\! \mod q, \, x+(c_i+1)t \!\!\! \mod q\}, \ \ \
i = 1,2,\ldots,\ell,
$$
for $x \in [0,q-1]$ and $t \in [1,q-1]$. In other words,
the edges of $H(q;\O)$ are precisely the proper solutions over $\Z_q$
of equations in $\O$. Therefore, a subset of $[0,q-1]$ contains no proper
solution over $\Z_q$ to any equation in $\O$ if
and only if it forms an independent set of vertices in $H(q;\O)$. Thus, any
lower bound on the independence number of $H(q;\O)$ is also a
lower bound on $s(q;\O)$. We will prove the bound of the theorem using
the following lower bound \cite[p.136]{Ber89} on
the independence number of a regular, $h$-uniform hypergraph, $H =
(V(H),E(H))$:
\begin{equation}
\alpha(H) \geq  \frac{|V(H)|}{|E(H)|^{1/h}}.
\label{indep_bound}
\end{equation}

It is easily seen that the hypergraph $H(q;\O)$ is 3-uniform and
$\ell(q-1)$-regular. Indeed, for any $x \in [0,q-1]$, $t \in
[1,q-1]$ and $c\in[1,q-2]$, the integers $x$, $x+t$ and $x+(c+1)t$
are distinct modulo $q$, since $q$ is a prime. Therefore, each edge
of $H(q;\O)$ contains exactly three distinct vertices, showing that
$H(q;\O)$ is 3-uniform. To see that the graph is
$\ell(q-1)$-regular, we only need to observe that for each vertex $x
\in [0,q-1]$, the triples
$$
\{x, \, x+t \!\!\! \mod q, \, x+(c_i+1)t \!\!\! \mod q\}, \ \ \
i = 1,2,\ldots,\ell,
$$
$1 \leq t \leq q-1$, form an exhaustive set of distinct hyperedges
containing the vertex $x$.

The number of edges in $H(q;\O)$ can be easily computed from the
fact that the graph is 3-uniform and $\ell(q-1)$-regular, so that we
must have $3|E(H(q;\O))|=\ell(q-1)|V(H(q;\O))|$. Consequently,
$|E(H(q;\O))|=(\ell/3)q(q-1)$. The theorem is proved by plugging
this into the bound of \eqref{indep_bound}. \endproof \mbox{ }

It is now only left to prove Theorem~\ref{Omega_theorem}. The proof
uses a technique due to F.A.\ Behrend
(see \cite[Section~4.3, Theorem~8]{GRS90}),
and hinges upon the following lemma.

\begin{lemma}
Given a system, $\Omega$, as in (\ref{Omega}),
let $D=\max_{1 \leq i \leq \ell}\,b_i$,
and let $q$ be an integer larger than $D$.
Pick an integer $n > 0$ such that $nD<q$,
and let $k = \lfloor(\log\,q)/\log(nD+1)\rfloor$.
Then, there exists a set, $S$, of integers from $[0,q-1]$, of cardinality
\begin{equation}\label{lobound}
|S| \geq \frac{(n+1)^{k-2}-1}k \notag
\end{equation}
such that $S$ does not contain a proper solution over $\Z_q$
to any of the equations in $\Omega$.
\label{Omega_lemma}
\end{lemma}

\begin{proof} Let $D$, $q$, $n$ and $k$ be as in the statement of the
lemma, and let $M = {(nD+1)}^k-1$.
For each $x \in [1,M]$, let $(x_0,x_1, \ldots, x_{k-1})$ be
the $(nD+1)$-ary representation of $x$, {\em i.e.},
$x=\sum_{i=0}^{k-1}\, x_i\, (nD+1)^i$ with $x_i \in [0,nD]$ for
$i = 0,1,\ldots,k-1$. We will refer to $(x_0,x_1, \ldots, x_{k-1})$
as the coordinate vector of $x$. Define
$$
    N(x)=\left(\sum_{0 \leq i \leq k-1}\, x_i^2\right)^{1/2}.
$$
In other words, $N(x)$ is the $l^2$-norm of the coordinate
vector $(x_0,x_1, \ldots, x_{k-1})$.
For an arbitrary integer $\rho \geq 1$, define the set
$$
R_{\rho,n}= \{{x\in [1,M]:\  0 \leq x_i \leq n\, \; \forall \, i,
\, N(x)^2 = \rho\}}.
$$
In other words, $R_{\rho,n}$ is the set of all integers $x \in [1,M]$
that satisfy two properties:
\begin{itemize}
\item[(i)] the digits $x_0, x_1, \ldots, x_{k-1}$
in the $(nD+1)$-ary expansion of $x$ all lie between 0 and $n$; and
\item[(ii)] $N(x)^2=\rho$, \emph{i.e.},
the coordinate vector of $x$ lies on the $l^2$-sphere of radius $\sqrt{\rho}$.
\end{itemize}

Our goal is to show that there exists a $\rho^* \geq 1$ such that
$R_{\rho^*,n}$ does not contain a proper solution
over $\Z_q$ to any equation in $\Omega$, and
$|R_{\rho^*,n}| \geq  \frac{(n+1)^{k-2}-1}k$. It then follows
that $R_{\rho*,n}-1 = \{u-1 : u \in R_{\rho*,n}\}$ is the
set $S$ in the statement of the theorem.

We will first show that for any $\rho \geq 1$,
$R_{\rho,n}$ cannot contain a proper solution
over $\Z_q$ to any equation in $\Omega$. In fact, it is enough to
show that $R_{\rho,n}$ cannot contain a proper \emph{integer}
solution to any equation in $\Omega$. This is because for any set
$\{u_1,u_2,\ldots,u_m,v\} \subset R_{\rho,n}$, we must have
\begin{eqnarray*}
\sum_{j=1}^m \, c_{i,j}\, u_j
&<& \sum_{j=1}^m c_{i,j} \left(\sum_{t=0}^{k-1} n {(nD+1)}^t\right) \\
&=& b_i \,\left(\sum_{t=0}^{k-1} n {(nD+1)}^t\right) \\
&\leq& (nD) \, \sum_{t=0}^{k-1} {(nD+1)}^t \\
&=& (nD+1)^k-1,
\end{eqnarray*}
and similarly,
$$
b_iv \leq (nD+1)^k-1,
$$
which together imply
\begin{eqnarray*}
|\sum_{j=1}^m \, c_{i,j}\, u_j - b_i v|
&<& \max\{\sum_{j=1}^m \, c_{i,j}\, u_j, \; b_i v\} \\
&\leq& (nD+1)^k-1 \ \ < \ \ q.
\end{eqnarray*}
Hence, $c_{i,j}\, u_j - b_i v \equiv 0 \pmod{q}$ if and only if
$c_{i,j}\, u_j - b_i v = 0$ (over the integers).

Note that, since $c_{i,j} \leq b_i \leq D$,
each digit $x_i$ in the $(nD+1)$-ary representation of
an element in $R_{\rho,n}$ is small enough so that there is no
carry over when performing any of the sums in $\Omega$. Hence,
adding numbers in $R_{\rho,n}$ corresponds to adding their coordinate
vectors. Now, suppose that the $i$th equation in
$\Omega$ has a solution $\{u_1,u_2,\ldots,u_m,v\} \subset
R_{\rho,n}$, {\em i.e.}, $\sum_{j=1}^m \, c_{i,j}\, u_j = b_i v$,
with $N(u_1)^2= N(u_2)^2 = \ldots = N(u_m)^2 = N(v)^2 = \rho$. Then,
$$v = \frac{1}{b_i} \sum_{j=1}^m \, c_{i,j}\, u_j,$$ which means
that the coordinate vector of $v$ is a convex combination of the
coordinate vectors of the $m$ integers $u_j$, $j = 1,\ldots,m$,
and all these vectors lie on the $l^2$-sphere of radius $\sqrt{\rho}$.
However, by the strict convexity of the $l^2$-norm, this can happen
only if all these coordinate vectors are identical, or equivalently,
only if $u_1 = u_2 = \ldots = u_m = v$. So,
$R_{\rho,n}$ cannot contain a proper integer solution to any
equation in $\Omega$.

At this point, the proof turns nonconstructive. Note that the union
$$
\bigcup_{\rho \geq 1} R_{\rho,n} =
\{x \in [1,M]:\ 0 \leq x_i \leq n \ \forall \, i\}
$$
contains $(n+1)^k-1$ points in all, since this is
the number of sequences $(x_0,\ldots,x_{k-1})$ such that
$0 \leq x_i \leq n$ for $i = 0,1,\ldots,k-1$.
Furthermore, for any $x \in \bigcup_{\rho \geq 1} R_{\rho,n}$,
we have $N(x)^2 = \sum_{i=0}^{k-1} x_i^2 \leq k \,n^2$, so that
$$
\bigcup_{\rho \geq 1} R_{\rho,n} = \bigcup_{\rho=1}^{kn^2} R_{\rho,n}.
$$
Thus, the union of the sets $R_{\rho,n}$, $\rho = 1,2,\ldots,kn^2$,
contains a total of $(n+1)^k-1$ points.
Hence, by the pigeon-hole principle, there exists a $\rho^* \in [1,kn^2]$
such that
\begin{equation}\label{set-size}
    |R_{\rho*,n}| \geq \frac{(n+1)^k-1}{k\,n^2} \geq
    \frac{(n+1)^{k-2}-1}k. \notag
\end{equation}
Finally, $S = R_{\rho*,n}-1 = \{u-1 : u \in R_{\rho*,n}\}$ is the
set whose existence is claimed in the statement of the theorem.
Indeed, $S \subset [0,M-1]$, and since $M = {(nD+1)}^k-1 \leq q$, we have
$S \subset [0,q-1]$. Moreover, as $(1,1,\ldots,1)$ is a solution to
every equation in $\O$, and $R_{\rho*,n}$
does not contain a proper solution to any equation in $\O$, $S$
cannot contain such a solution either. \end{proof} \mbox{}

We can now give the proof of Theorem~\ref{Omega_theorem}.

\emph{Proof of Theorem~\ref{Omega_theorem}\/}:
Given $q > D^2$, pick an arbitrary $\epsilon > \log D/ \log q$.
Then, choosing $n=\lfloor\frac{1}{D}\,q^{\epsilon}\rfloor$
and applying Lemma~\ref{Omega_lemma},
we get the following lower bound on $s(q;\O)$:
\begin{equation}
s(q;\O) \geq \epsilon \, D^{2-\frac{1}{\epsilon}} \,
q^{1-2\epsilon}(1+o(1)),
\label{s_bnd}
\end{equation}
where $o(1)$ denotes a correction factor that goes to zero as
$q \rightarrow \infty$.

Now, it may be verified that the value of $\epsilon$
that maximizes the function $f(\epsilon) =
\epsilon \, D^{2-\frac{1}{\epsilon}} \, q^{1-2\epsilon}$ is
$\left(1 + \sqrt{1 + 8 (\log D)(\log q)}\right)/(4 \log q)
\approx \sqrt{\frac{1}{2} \log D / \log q}$. For $q > D^2$,
$\sqrt{\frac{1}{2} \log D / \log q} > \log D/ \log q$,
so the bound in \eqref{s_bnd} applies with
$\epsilon = \sqrt{\frac{1}{2} \log D / \log q}$. Plugging this
value of $\epsilon$ into \eqref{s_bnd}
and manipulating the resulting expression,
we obtain the bound of the theorem. \endproof\mbox{ }

\section*{Acknowledgment} The authors are grateful to the anonymous
reviewers for their suggestions which led to a significantly improved
presentation of the results in the paper.
The authors would also like to thank Daniel Bennett and Stefan Laendner
for helping with some of the simulations.

\end{document}